\documentclass[twocolumngrid,prd,superscriptsize,reprint]{revtex4-1}
\usepackage[]{graphicx}
\usepackage{dcolumn}
\usepackage{bm}

\usepackage{amsmath,amssymb,amsfonts}
\usepackage{fancyhdr}
\usepackage{lipsum}
\usepackage{subcaption}
\usepackage{multirow}
 \usepackage[colorlinks]{hyperref}
\hypersetup{
     colorlinks = false,
   }

\newcommand{\comment}[1]{}

\usepackage{color}
\usepackage{babel}
\newcommand{\bea}{\begin{eqnarray}}
\newcommand{\eea}{\end{eqnarray}}

\newcommand{\pa}{\partial}

\newcommand{\be}{\begin{equation}}
\newcommand{\ee}{\end{equation}}
\numberwithin{equation}{section}

\begin{document}

\title[]{Relativistic Moduli Space for Kink Collisions}

\author{C. Adam}
\email[]{adam@fpaxp1.usc.es}
\affiliation{Departamento de F\'isica de Part\'iculas, Universidad de
Santiago de Compostela and \\
Instituto Galego de F\'isica de Altas Enerxias (IGFAE), E-15782
Santiago de Compostela, Spain}

\author{N.~S. Manton}
\email[]{N.S.Manton@damtp.cam.ac.uk}
\affiliation{Department of Applied Mathematics and Theoretical Physics,
University of Cambridge,
Wilberforce Road, Cambridge CB3 0WA, U.K.}

\author{K. Oles}
\email[]{katarzyna.slawinska@uj.edu.pl}
\author{T. Romanczukiewicz}
\email[]{tomasz.romanczukiewicz@uj.edu.pl}
\author{A. Wereszczynski}
\email[]{andrzej.wereszczynski@uj.edu.pl}
\affiliation{Institute of Theoretical Physics, Jagiellonian University,
Lojasiewicza 11, Krak\'{o}w, Poland}

\begin{abstract}
The moduli space approximation to kink dynamics permits a relativistic
generalization if the Derrick scaling parameter is used as a
collective coordinate. We develop a perturbative approach to the
resulting relativistic moduli space by expanding the Derrick scaling
parameter about unity and treating the higher-order Derrick
modes as new degrees of freedom. This approach allows us to resolve
(coordinate) singularities order-by-order, and
systematically incorporates relativistic corrections {\it perturbatively}
in kink scattering. It gives an excellent description of
kink-antikink collisions in $\phi^4$ field theory already at first
order, and at higher order, reproduces the fractal structure in the
formation of the final state with an error of only $4\%$. 

\end{abstract}
\maketitle

\section{Introduction}

Topological solitons \cite{R, SM, Shnir} are solutions of nonlinear
field equations possessing, at first glance, two quite opposed
features. On the one hand, they
are particle-like objects, whose energy density is localized in
a certain region of space. On the other, they carry a topological
charge, a quantity characterizing a solution globally, that depends
on the field behavior at infinity. Partly owing to this juxtaposition
of short-range and long-range features, the dynamics of
topological solitons is quite involved, and leads to complex
patterns of final states in scattering processes.
Except in the rare cases of integrable theories, many aspects of
soliton scattering are still far from being fully
understood. Among these is the fractal velocity-dependence of the
final state in kink-antikink collisions, associated with
the resonant coupling of translational motion to oscillatory modes
\cite{Sug, CSW, CPS, mal, gani, Good1, Good2,
Weigel, AI2, Kev, long, Sim, AI1, Aza, luchini, MORW, Vill}, which can
be normal modes or quasi-normal modes \cite{DTr-q} hosted by free
solitons, or the internal modes hosted by ephemeral configurations
occurring during the collision \cite{DTr, sphaleron}.
Also intriguing is the recently discovered spectral wall phenomenon
\cite{spectral-wall, sw2}, caused by the transition of a normal
mode into the continuum spectrum as solitons approach each other.
Finally, there is the famous, long-standing problem of the
soliton resolution conjecture \cite{resol1, resol2}.

One method to reduce the complexity of topological soliton dynamics
is to construct a collective coordinate model (CCM). CCM dynamics is
sometimes referred to as moduli space dynamics. In
this approximation, a field theory Lagrangian $L[\phi]$ that 
incorporates infinitely many field degrees of freedom
is truncated to a dynamical system $L[{\bf X}]$ with finitely many
collective coordinates $X^i$, $i=1,\dots,N$, also called moduli.
Despite this rather drastic simplification, under certain
circumstances the CCM accurately describes the full soliton dynamics,
because only a subset of field configurations play an
important role, while the rest may be neglected.

The moduli space manifold $\mathcal{M}$ offers a more global
perspective than the local collective coordinates $X^i$ defined on
it. In the cases we consider, $\mathcal{M}$ has a Riemannian metric
inherited from the kinetic term of the field theory Lagrangian, and
a potential inherited from the remaining terms including the
gradient term. As we will see below, with the obvious choice of
collective coordinates for interacting solitons, the metric on
$\mathcal{M}$ sometimes has a singularity, which needs to be removed.
This is not always possible by simply changing coordinates. 

For solitons that have no static interactions (solitons of Bogolmol'nyi, or BPS, type)
\cite{Bo, JT, Ma8, nick, AH, Sam, solv} there is a canonical moduli
space. But in general, there is no canonical way to construct the
restricted set of field configurations in the moduli space, and
one needs to make an educated guess. This is the case for
weakly interacting solitons \cite{unstable, Sp}, and especially for
processes like kink-antikink (KAK) collisions in $\phi^4$ field
theory, whose configurations are far from being BPS \cite{MORW}.

Very frequently, the collective coordinates are chosen in such a way
that the resulting CCM is nonrelativistic, even though 
the original field theory is Lorentz invariant. This is rather
undesirable, both theoretically and from the
point of view of applications. There are
processes in which the relativistic nature of the colliding solitons
is crucial for explaining the observed data. See,
e.g., the role of Lorentz contractions in KAK collisions in
$\phi^4$ theory \cite{MORW}. For the case of kink dynamics, this
shortcoming can be partially resolved by including the Derrick scaling 
deformation, which allows for a Lorentz contraction of a single kink,
as was originally observed in \cite{Rice}. The result is a
relativistic CCM, although it is still not fully relativistic, as it
does not include radiation modes.

In multi-kink sectors, inclusion of the Derrick deformation can lead to the
appearance of singularities in the relativistic CCM. It was
claimed earlier that these singularities can be removed by a rather
nontrivial change of the collective coordinates \cite{caputo}. This
issue is reinvestigated here, and it is shown that the singularity appears to
be essential. To circumvent this, we present a novel approach to the
relativistic CCM. We develop a perturbative expansion
where relativistic corrections are taken into account order-by-order
in the squared collision velocity, leading to a sequence of
higher-order Derrick modes. A simple redefinition of the amplitudes of
these Derrick modes then permits a resolution of the singularities in
the description of kink-antikink collisions. This elevates the
perturbative relativistic collective coordinate model (pRCCM) to an
accurate {\it quantitative} tool for
understanding the dynamics of topological solitons. 

\section{Derrick Deformation and Relativistic Moduli Space}

Consider a real scalar- or vector-valued field $\phi$ in $(1+1)$-dimensions
(with all internal indices suppressed), governed by the Lorentz-invariant
Lagrangian
\bea
L[\phi] &=& \int_{-\infty}^\infty  \mathcal{L} [\phi] \, dx \nonumber \\
&=& \int_{-\infty}^\infty  \left(
\frac{1}{2}(\pa_t\phi)^2 - \frac{1}{2}(\pa_x\phi)^2 - U(\phi)\right) dx
\label{model}
\eea
that combines the standard space-time derivative terms with a non-negative
potential $U$.  Now suppose we have a restricted set of static field
configurations -- a moduli space of configurations -- capturing the
main features of a solitonic process at each instant of time,
\be
\mathcal{M}=\{ \Phi(x;X^i),\;  i=1, \dots, N \}.
\ee
In a BPS theory, there is a {\em canonical} moduli space of static
multi-soliton solutions with equal energies.
There is also a canonical moduli space for a
single soliton in a non-BPS theory. Multi-soliton configurations in a
non-BPS theory are not so easily defined, but a linear
superposition of single solitons and/or antisolitons is usually
possible. Frequently, further moduli parametrizing fundamental soliton
excitations should be included, because these are excited in soliton
collisions. These moduli can be chosen to be the amplitudes of
positive-frequency modes, i.e., {\it normal modes} or
even {\it quasi-normal modes}, hosted by the solitons. However,
we will make another choice later.  

In the CCM, all the moduli of the interacting solitons are
promoted to time-dependent variables, $X^i(t)$. Then, inserting the
restricted field configurations $ \Phi(x; X^i(t))$ into the Lagrangian
density $\mathcal{L}$ and performing the spatial integrations, we arrive
at an effective Lagrangian for motion on the $N$-dimensional moduli space
$\mathcal{M}$. This can be interpreted as a mechanical model for
interacting solitons with $N$ degrees of freedom. The
Lagrangian on $\mathcal{M}$ is
\be
L[{\bf X}]=\int_{-\infty}^\infty  \mathcal{L}[\Phi(x; X^i(t))] \, dx
= \frac{1}{2} g_{ij}({\bf X}) \dot{X}^i \dot{X}^j - V({\bf X}) \,,
\label{eff-lag}
\ee
where
\be
g_{ij}({\bf X})=\int_{-\infty}^\infty \frac{\partial \Phi}
{\partial X^i} \frac{\partial \Phi}{\partial X^j} \, dx
\label{modmetric}
\ee
is the metric on $\mathcal{M}$, while
\be
V({\bf X})=\int_{-\infty}^\infty \left( \frac{1}{2}
\left( \frac{\partial \Phi}{\partial x}
\right)^2 + U(\Phi) \right) \, dx
\label{modpot}
\ee
is the potential energy. Ideally, the metric
should not possess any singularities except where $V$
diverges, which is naturally the boundary of $\mathcal{M}$, and of
course unattainable in any finite-energy field evolution.

However, there are examples in which the moduli space metric is
singular at points where $V$ is finite. Note that a metric is
singular, not only where a metric component diverges, but also where
the metric tensor as a whole degenerates (i.e., is not positive definite)
and its determinant vanishes. Sometimes the singularity is
resolved by a better choice of coordinates, and we will encounter
examples of this. In other cases, the singularity is
physical, for example, corresponding to the
location of a spectral wall, a barrier in the soliton dynamics caused
by the transition of a normal mode into the continuum spectrum.

The simplest moduli space is that of a single kink
$\Phi_K(x)$, a solution of the static BPS equation
\be
\frac{d\phi}{dx} = \sqrt{2U(\phi)} \,.
\label{BPSkink}
\ee
From (\ref{modpot}) we see that the energy (mass) of the static kink
is
\be
M = \int_{-\infty}^\infty
\left( \frac{d\Phi_K(x)}{dx} \right)^2 \, dx \,.
\label{kinkmass}
\ee
We assume throughout that the kink interpolates between isolated vacua
of the potential $U$ and that $U$ is symmetric between these
vacua. The kink shape is then reflection-antisymmetric. By translation
symmetry, the kink centre is an arbitrary real constant $a$, leading
to a one-parameter family of equal-energy solutions
$\mathcal{M}=\{\Phi_K(x-a)\}$. This canonical moduli space is
the infinite line $\mathbb{R}$ with coordinate $a$.

Note that the derivative of $\Phi_K(x-a)$ w.r.t. $a$ is minus the
derivative w.r.t. $x$, and that this is a reflection-symmetric
function. The metric on $\mathcal{M}$ is therefore a
constant equal to the kink mass $M$,
\be
g_{aa} = \int_{-\infty}^\infty \left( \frac{\partial \Phi_K }{\partial a}
  (x-a) \right)^2 dx=M \,,
\ee
because the integral here is a translate of (\ref{kinkmass}). The
potential on $\mathcal{M}$ has the same value,
\bea
V(a) &=& \int_{-\infty}^\infty \left( \frac{1}{2} \left(
\frac{\partial \Phi_K}{\partial x} (x-a)\right)^2 +U(\Phi_K(x-a))
\right)dx \nonumber \\
&=& M \,.
\eea
The CCM therefore has the Lagrangian 
\be
L[a]=\frac{1}{2} M \dot{a}^2 - M \,,
\ee
modelling a non-relativistic point particle in a constant,
immaterial potential. Solutions of the equation of motion
$\ddot a = 0$ model kink motion at arbitrary constant velocity
$\dot{a}=v$. Importantly, the Lorentz invariance of the original field
theory is lost, and this feature is inherited by the more complicated
dynamics in, for example, kink-antikink collisions. This is because the
simplest CCM describing such collisions is constructed using a
superposition of static kinks and antikinks.

Typically, an improved soliton moduli space is constructed by including
deformations arising from the linear perturbations of the static
solitons, i.e., positive-frequency normal
modes of the second variation operator, known as shape modes.

There is, however, another physically well motivated deformation,
the {\it  Derrick} or {\it scaling deformation}, which in one
spatial dimension is simply $x \mapsto bx$. When the Derrick modulus,
or scaling parameter, $b$ is included in the single-kink
sector, the set of configurations is \cite{Rice}.
\be
\mathcal{M}=\{\Phi_K(b(x-a))\} \,.
\ee
$a$ still defines the position of the kink and therefore
$a\in \mathbb{R}$. On the other hand, $b > 0$, because
when $b=0$ the configuration fails to satisfy the kink boundary
conditions, and when $b<0$ the kink becomes an antikink. For $b$
close to unity, the linearized deformation of the kink centred at the
origin is
\be
\eta_D(x) = x \frac{d\Phi_K(x)}{dx} \,,
\ee
which we call the {\it Derrick mode} of the kink. The Derrick mode
is not generally related to a shape mode, but one of its advantages
is that it can be used in any model with kinks, whether or not the
kink has a shape mode. 

The moduli space $\mathcal{M}$ is now two-dimensional
and has the diagonal metric
\bea
g_{aa} &=& \int_{-\infty}^\infty b^2 \Phi^{'2}_K (b(x-a)) \, dx =
M b \,, \nonumber \\
g_{bb} &=& \int_{-\infty}^\infty (x-a)^2 \Phi^{'2}_K (b(x-a)) \, dx =
\frac{Q}{b^3} \,.
\eea
Here, $\Phi'_K$ is the derivative of $\Phi_K$, and therefore the kink's
translation zero mode, $M$ is the kink mass (\ref{kinkmass}), while
the constant $Q$ is the second moment of the static kink's energy density, 
\be
Q=\int_{-\infty}^\infty x^2 \left( \frac{d \Phi_K(x)}
{dx}\right)^2 dx \,.
\ee
$g_{ab} = 0$ because the $a$-derivative and $b$-derivative of the kink
have opposite symmetries when $U$ has the reflection symmetry assumed earlier.

$\mathcal{M}$ is an anticlastic surface since it has Ricci curvature 
\be
R=-\frac{b}{Q} \,,
\ee
which is everywhere negative and, up to a multiplicative constant,
just the Derrick modulus $b$.  

The potential on $\mathcal{M}$ has the simple form
\be
V(b) = \frac{1}{2} M \left( b + \frac{1}{b} \right) \,,
\ee
so the CCM Lagrangian is
\be
L[a,b] = \frac{1}{2} M b \dot{a}^2 + \frac{1}{2} \frac{Q}{b^3} \dot{b}^2
- \frac{1}{2} M \left( b + \frac{1}{b} \right) \,.
\label{CCMkinkLag}
\ee
The resulting equations of motion on $\mathcal{M}$ are
\bea
\hspace*{-0.8cm} && \frac{d}{dt} (b\dot{a}) = 0 \,, \\
\hspace*{-0.8cm} && Q \frac{d}{dt} \left(\frac{\dot{b}}{b^3} \right) +
\frac{3Q}{2} \frac{\dot{b}^2}{b^4}  +
\frac{M}{2} \left(1 - \frac{1}{b^2} -\dot{a}^2 \right) = 0 \,,
\eea
and these can be integrated once to give 
\bea
\hspace*{-0.8cm} && Mb\dot{a} = P \,, \label{Pcons} \\
\hspace*{-0.8cm} && \frac{1}{2} M b \dot{a}^2 + \frac{1}{2}
\frac{Q}{b^3} \dot{b}^2 + \frac{1}{2} M \left( b + \frac{1}{b} \right) = E \,,
\label{Econs}
\eea
where $P$ and $E$ are the conserved momentum and energy. 

Importantly, there exist stationary, non-oscillating solutions with
\be
\dot{a}=v \,, \quad b= \gamma \equiv \frac{1}{\sqrt{1-v^2}} \,,
\label{statsoln}
\ee
so $\Phi(x,t) = \Phi_K(\gamma(x-vt))$.
For these solutions, $P$ and $E$ obey the relativistic relation
$E^2 - P^2 = M^2$, as originally observed in \cite{Rice}. This is
the crucial result of this section. The inclusion of the Derrick
deformation allows a moving kink to Lorentz
contract. In other words, it preserves the relativistic invariance of
the field theory at the level of the moduli space dynamics. Note that
the solutions in the moduli space are along lines characterized by the
Ricci curvature being a constant proportional to the Lorentz
contraction factor $\gamma$ of the kink.

We will call this CCM the {\it relativistic collective coordinate model},
and the underlying moduli space $\mathcal{M}$ with coordinates
$(a,b)$ the {\it relativistic moduli space} for a single kink. In
fact, we could eliminate the modulus $b$ from the CCM using the
stationary solution $b=1/\sqrt{1-\dot{a}^2}$, where $\dot{b}=0$.
Then we find
\be
L[a]= -M\sqrt{1-\dot{a}^2} \,.
\ee
This is precisely the Lagrangian of a {\it relativistic point particle}
with rest mass $M$. Hence, we explicitly obtain a relativistic
generalization of the non-relativistic CCM for a single kink.

Rather remarkably, the general oscillatory dynamics on $\mathcal{M}$
reduces to simple harmonic motion, for any momentum $P$ and energy
$E$ \cite{Rice}. Let us see how this comes about in the case $P = 0$.
Here $\dot{a} = 0$ so the energy conservation equation (\ref{Econs})
simplifies to
\be
\frac{1}{2} \frac{Q}{b^3} \dot{b}^2  
+ \frac{1}{2} M \left( b + \frac{1}{b} \right) = E \,,
\ee
which after the change of variable $\ell = 1/b$ becomes
\be
\dot{\ell}^2 + \frac{M}{Q}\left( \ell - \frac{E}{M} \right)^2 =
\frac{M}{Q}\left( \frac{E^2}{M^2} - 1 \right) \,.
\ee
For any $E \ge M$ this is the first integral of a simple harmonic
oscillator with frequency $\omega^2 = M/Q$. Because of the shift of
the centre of oscillation to $\ell = E/M$, the motion is
restricted to the range $\ell > 0$. The linearized Derrick mode
of a kink, $\eta_D$, has the same frequency $\omega_D^2 = M/Q$, with
centre of oscillation at $\ell = 1$. 

\section{Derrick Deformation in the Kink-Antikink Sector}

To model kink-antikink collisions, we consider the field configurations
obtained by the symmetric superposition
\be
\Phi_{KAK}(x;a,b) = \Phi_K(b(x+a)) - \Phi_K(b(x-a)) + \Phi_{vac}
\label{KAKConfigs}
\ee
with time-dependent $a$ and $b$. Here, $\Phi_K(x)$ is the basic kink
solution and $\Phi_{vac}$ the constant needed to satisfy the vacuum
boundary conditions. The kink and antikink are at $-a$ and $a$
respectively, and the Derrick modulus $b$ has the
same value for both, ensuring the symmetry. The
configurations (\ref{KAKConfigs}) are unchanged if $a \mapsto -a$ and
$b \mapsto -b$, so we may assume that $b > 0$ and $-\infty < a < \infty$.

The moduli space is two-dimensional, with metric components $g_{aa}$,
$g_{ab}$ and $g_{bb}$ given by the general integral formula
(\ref{modmetric}). These metric components satisfy an interesting
identity that is actually valid for kink-kink and
kink-antikink superpositions of the general form 
\be
\Phi(x;a,b) = \Phi_K(b(x+a)) \pm \Phi_K(b(x-a)) + \Phi_0
\ee
where $\Phi_0$ is any constant. The proof starts with the derivatives
of these configurations
\bea
\partial_a\Phi &=& b\Phi'_{K+} \mp b\Phi'_{K-} \,, \nonumber \\
\partial_b\Phi &=& (x+a)\Phi'_{K+} \pm (x-a)\Phi'_{K-} \,, 
\eea
where 
\be
\Phi'_{K\pm} = \left. \frac{d\Phi_K(y)}{dy}
\right|_{y=b(x\pm a)} \,.
\ee
The mixed metric component is therefore 
\bea
&g_{ab}&=\int_{-\infty}^\infty (\partial_a \Phi) (\partial_b \Phi)
\, dx \nonumber \\
&=& \int_{-\infty}^\infty \bigl( b(x+a) \Phi^{'2}_{K+}
  - b(x-a) \Phi^{'2}_{K-} \mp 2ab \, \Phi'_{K+}\Phi'_{K-} \bigr) \, dx \nonumber \\
&=& \mp 2ab \int_{-\infty}^\infty \Phi'_{K+}\Phi'_{K-}  \, dx \,,
\label{Agab}
\eea
where the last equality follows from the vanishing of the first two
integrals due to the kink's reflection-antisymmetry, and the diagonal
component $g_{aa}$ is 
\bea
g_{aa}&=&\int_{-\infty}^\infty (\partial_a \Phi)^2 \, dx \nonumber \\
&=& b^2 \int_{-\infty}^\infty \left( \Phi^{'2}_{K+} + \Phi^{'2}_{K-}
  \mp 2 \Phi'_{K+}\Phi'_{K-} \right) \, dx \nonumber \\
&=& 2bM \mp 2 b^2  \int_{-\infty}^\infty \Phi'_{K+}\Phi_{K-} \, dx \,,
\label{Agaa}
\eea
where $M$ is the kink mass. ($g_{bb}$ is not needed here.)
Comparing (\ref{Agab}) and (\ref{Agaa}) we obtain the identity
\be
ag_{aa} - bg_{ab} = 2Mab \,.
\label{Identityab}
\ee
It is sometimes useful to reparametrize the kink-antikink
configurations (\ref{KAKConfigs}) as
\be
\Phi_{KAK}(x;b,c) = \Phi_K(bx+c) - \Phi_K(bx-c) + \Phi_{vac}
\label{KAKbc}
\ee
where $c = ba$ and $b > 0$, $-\infty < c < \infty$. The metric
components, now denoted ${\tilde g}_{bb}$, ${\tilde g}_{bc}$ and
${\tilde g}_{cc}$, satisfy the simpler identity
\be
b^2{\tilde g}_{bc} = -2Mc \,.
\label{Identitybc}
\ee
and the same identity holds in the kink-kink case.
These identities will be verified in examples occurring below.

It is important for us to investigate whether the moduli space of
kink-antikink configurations (\ref{KAKbc}) is globally smooth.
For $c > 0$, the derivatives of $\Phi_{KAK}$ w.r.t. $b$ and $c$ are non-zero
and independent, so this subregion of moduli space is
smooth. Similarly, the region $c < 0$ is smooth. However, $c=0$ gives
the vacuum configuration for any $b$, so it has zero derivative w.r.t. $b$.
The two smooth parts of the moduli space are therefore glued at a
single point, giving a total space with a singularity, somewhat like a
double-cone. No change of coordinates removes this singularity.

We can find the metric and its curvature in the neighbourhood of $c=0$
more precisely by expanding to linear order in $c$. At this order the
configurations (\ref{KAKbc}) become
\be
\Phi_{KAK}(x;b,c) = 2c\Phi'_K(bx)
\ee
(dropping the constant $\Phi_{vac}$). The derivatives needed for
the metric are
\be
\frac{\pa \Phi_{KAK}}{\pa b} = 2cx \Phi''_K(bx) \,, \quad
\frac{\pa \Phi_{KAK}}{\pa c} = 2 \Phi'_K(bx) \,.
\ee
The metric components on the moduli space are therefore
\bea
{\tilde g}_{bb} &=& 4c^2 \int_{-\infty}^\infty x^2
\Phi^{''2}_K(bx) \, dx \,, \nonumber \\
{\tilde g}_{bc} &=& 4c \int_{-\infty}^\infty
x\Phi''_K(bx)\Phi'_K(bx) \, dx = -\frac{2c}{b^2} M \,,
\nonumber \\
{\tilde g}_{cc} &=& 4 \int_{-\infty}^\infty
\Phi^{'2}_K(bx) \, dx = \frac{4}{b} M \,.
\eea
The last result follows from (\ref{kinkmass}), and the middle result
is obtained by integrating by parts or using the identity
(\ref{Identitybc}). The complete metric for small $c$ is therefore
\be
ds^2 = 4M \left( \frac{Sc^2}{b^3} db^2 - \frac{c}{b^2} dbdc
+ \frac{1}{b} dc^2 \right) \,,
\label{metricsmallc}
\ee
where
\be
S = \frac{\int_{-\infty}^\infty y^2 \Phi^{''2}_K(y) \, dy}
{\int_{-\infty}^\infty \Phi^{'2}_K(y) \, dy} \,.
\ee

For $c > 0$ it is helpful once more to change variables, to a
variant of plane polar coordinates. Let $\sigma = c/\sqrt{b}$ and
$\tau = \log(b)$, where $\sigma > 0$ and $-\infty < \tau < \infty$.
The metric (\ref{metricsmallc}) becomes
\be
ds^2 = 4M \left( d\sigma^2
+ \left(S - \frac{1}{4}\right)\sigma^2 d\tau^2 \right) \,.
\ee
This is a multiple of the standard flat metric on the plane,
but the origin is removed and the angle $\tau$ has infinite
range, so the surface is the infinite-sheeted universal cover of the
punctured plane. By symmetry, the surface for $c < 0$ is geometrically
similar. These two smooth surfaces are glued together at the single
point $\sigma = 0$, corresponding to the vacuum configuration $c=0$.
The total moduli space (for small $c$) is therefore an
infinite-sheeted version of a flat double-cone, and is singular.

Caputo {\it et al.} \cite{caputo} carried out a similar calculation,
which they applied to both sine-Gordon (sG) and $\phi^4$-theory kink-antikink
pairs. They also found a smooth surface by restricting the sign of one
of the moduli and changing variables, and then stated without detailed
justification that the complete moduli space was smooth, because the
change of variables could be extended to all values of the relevant
moduli. However, the calculation above indicates that a
singularity is unavoidable.

Our conclusion is that the kink-antikink moduli space constructed by
direct superposition of kink and antikink configurations, parametrized
by their position and Derrick moduli, is not smooth. This moduli space
also suffers from a more practical difficulty. The field
configurations with $c > 0$ have values everywhere greater than
$\Phi_{vac}$, and similarly those with $c < 0$ have values everywhere less than
$\Phi_{vac}$. The configurations that occur instantaneously in true field
theory simulations of kink-antikink scattering do not generally have
this property. The sG model is an exception. Here, the exact
kink-antikink scattering and breather solutions pass from one side of
$\Phi_{vac}$ to the other at one instant, for all $x$. Because
of this, sG solutions can be modelled well
using the moduli space we have been considering. However, in $\phi^4$
theory, and probably generically in non-integrable field theories,
the instantaneous kink-antikink field configurations often take
values on both sides of $\Phi_{vac}$, depending on $x$. Such dynamics
requires modelling with a different set of configurations and with
a different type of moduli space, and in the next section we introduce
our proposed perturbative framework to deal with this.  

\section{Perturbative Relativistic Moduli Space}

Here we present a novel and simple resolution of the singularity at
$a=0$ (or equivalently $c=0$) for KAK collisions, when $b$ is close to
1, i.e., $b=1+\epsilon$, with $\epsilon$ small. This
includes the physically relevant regime $b \gtrsim 1$,
corresponding to incoming or outgoing solitons with not too high
but still relativistic velocities. For free solitons
$b=\gamma = 1+ \frac{1}{2}v^2+ \frac{3}{8}v^4+o(v^4)$. Therefore,
$\epsilon = \frac{1}{2}v^2 + o(v^2)$, so the expansion in $\epsilon$
is just an expansion in $v^2$. In other words, relativistic corrections
are included {\it perturbatively}. This approach can be applied to all
kink models. In addition, it introduces in a natural way an arbitrary
number of new moduli.

\subsection{Single-kink sector}

Let us begin with the single kink, and set $b=1+\epsilon$. The
restricted set of configurations is a truncation of the Taylor expansion
\bea
\hspace*{-0.8cm} && \Phi_K(b(x-a)) = \Phi_K((1+\epsilon)(x-a))  \nonumber \\
&& \quad\quad = \sum_{k=0}^n \frac{\epsilon^k}{k!}(x-a)^k \Phi_K^{(k)}(x-a)
+o(\epsilon^n) \,,
\eea
where $\Phi_K^{(k)}$ denotes the $k$-th derivative of the kink.
A key idea is that we now replace the sequence of powers of $\epsilon$ by
{\it new, independent} moduli, denoted $C_1,\dots,C_n$, leading to the
following set of configurations,
\bea
\Phi_K(x;a,{\bf C}) &=& \Phi_K(x-a) \nonumber \\
&+& \sum_{k=1}^n \frac{C_{k}}{k!}(x-a)^k\Phi_K^{(k)}(x-a) \,.
\label{K-pert-gen}
\eea
The original two-dimensional relativistic CCM has been replaced by an
$(n+1)$-dimensional {\it perturbative} relativistic CCM (pRCCM),
where $n$ is the order of the expansion. Effectively, the $k$-th term
in the expansion introduces its own $k$-th order {\it Derrick mode}.
Note that the first-order Derrick mode is the
same as the Derrick mode $\eta_D$ that we introduced previously.

\subsection{Kink-antikink sector}

In the KAK sector, we can now resolve the singularity at $a=0$.
As usual, we start with simple superpositions of kink and antikink
solutions located at $-a$ and $a$, respectively. However, for the
kink and antikink we assume the truncated expansion (\ref{K-pert-gen}), so
\bea
 &\vspace*{-1.0cm}&\vspace*{-1.0cm} \Phi_{KAK}(x;a,{\bf C}) = (\Phi_K(x+a)-\Phi_K(x-a)) +\Phi_{vac} \nonumber
 \\
&& + \sum_{k=1}^n \frac{C_{k}}{k!} \bigl((x+a)^k\Phi_K^{(k)}(x+a)
- (x-a)^k\Phi_K^{(k)}(x-a)\bigr) \,. \nonumber  \\
&&
\label{pKAK-gen}
\eea
As $a\to 0$ the terms multiplied by $C_k$ vanish linearly, as 
\bea
&& (x+a)^k \Phi_K^{(k)}(x+a) - (x-a)^k\Phi_K^{(k)} (x-a) \nonumber \\
&& \quad = 2\left(kx^{k-1}\Phi_K^{(k)}(x)+  x^k \Phi_K^{(k+1)}(x) \right) a
+ o(a) \,. \nonumber \\
&&
\eea
This produces a null vector problem for the moduli $C_k$, but one that
can be easily resolved by our standard technique \cite{MORW}. Namely,
we make the replacement
\be
C_{k} \mapsto \frac{C_{k}}{\tanh(a)} \,.
\ee
Because $\tanh(a)$ is linear in $a$ for small $a$,
this replacement applied to (\ref{pKAK-gen}) gives a
set of smooth configurations leading to a smooth, finite metric
and potential for any $a,C_1,\dots,C_n$.

\section{Sine-Gordon model}

In this section we illustrate the application of the relativistic
collective coordinate model, in a case where it works well.
Two-soliton dynamics in the sine-Gordon (sG) model is the simplest case of
interacting solitons, and as the theory is integrable, we can compare
the  relativistic CCM with the exact solutions. We also investigate
the effect of the perturbative modification (pRCCM). 

The sG model, with Lagrangian
\be
L_{sG}[\phi]=\int_{-\infty}^\infty \left( \frac{1}{2}(\pa_t\phi)^2
  -\frac{1}{2}(\pa_x\phi)^2 - (1-\cos \phi)  \right) \, dx \,, \label{K-sG}
\ee
possesses the well-known static kink
\be
\Phi_K(x;a) = 4\arctan e^{x-a} \,,
\ee
with mass $M=8$. The non-relativistic single-kink CCM uses these
configurations parametrized by a dynamical position $a(t)$,
and has Lagrangian
\be
L[a]= 4\dot{a}^2 - 8 \,.
\ee

To obtain the relativistic version, we include the Derrick deformation.
This leads to the Lagrangian of the form (\ref{CCMkinkLag}),
\be
L[a,b]=4b \dot{a}^2 + \frac{\pi^2}{3b^3} \dot{b}^2
- 4 \left(b +\frac{1}{b} \right) \,, \label{K-sG-RCCM}
\ee
whose equations of motion have the solution (\ref{statsoln}),
modelling a stationary, Lorentz-contracted kink.

Note that the sG kink does not host any positive-frequency normal
modes. Despite this, the Derrick mode still exists and has frequency
$\omega_D^2= 12/\pi^2 \approx 1.21585$, which is above the continuum
threshold $\omega^2=1$.

\subsection{Kink-kink solution}

The exact kink-kink (KK) solution of the sG model is 
\be
\Phi_{KK}(x,t)=4\arctan \frac{v \sinh  (\gamma x) }{ \cosh (\gamma vt)} \,.
\label{exact-kk}
\ee
To show that this solution can be obtained within the relativistic
collective coordinate framework, we consider the moduli space
$\mathcal{M}$ of symmetric superpositions of kinks,
\bea
\hspace*{-0.3cm} \Phi_{KK}(x;a,b) \hspace*{-0.0cm} &=&
\hspace*{-0.0cm}  4\arctan e^{b(x+a)} + 4 \arctan e^{b(x-a)} -2\pi
\nonumber  \\
&=&\hspace*{-0.0cm}  4\arctan \frac{\sinh bx}{\cosh ba} \,,
\label{eff-kk}
\eea
with $b>0$. Because the kinks are identical, these configurations
are invariant under $a \mapsto -a$. This means that the
modulus $a$ can be assumed to be nonnegative, $a\geq
0$. Hence the kinks cannot cross over, and at closest approach they are
on top of each other. As we will see below, the resulting moduli
space $\mathcal{M}$ is incomplete and has a boundary at $a=0$, reachable
after a finite time. This problem is resolved by extending the set
of configurations (\ref{eff-kk}), similarly as in ref.\cite{zero-vec}.
\begin{figure}
 \includegraphics[width=1.0\columnwidth]{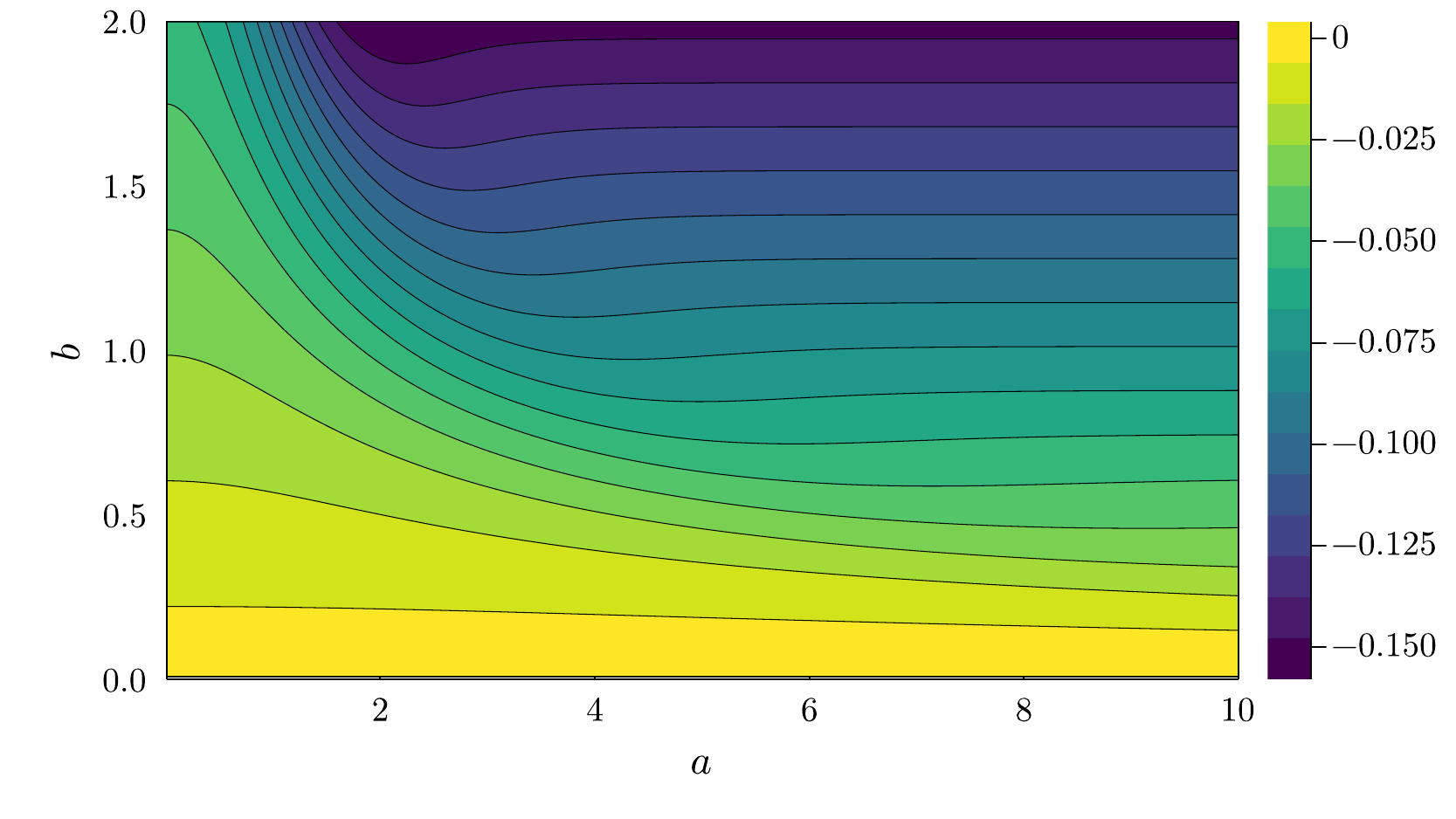}
 \includegraphics[width=1.0\columnwidth]{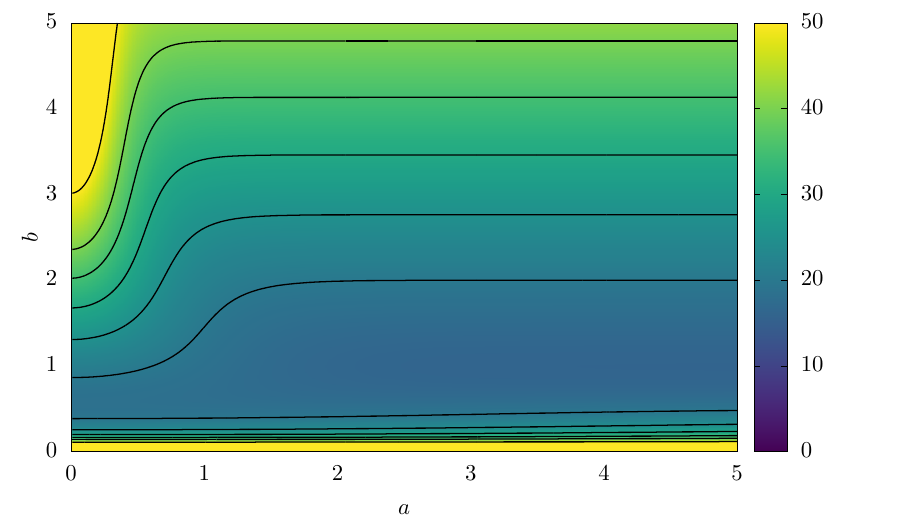}
 \caption{The Ricci curvature $R$ (upper panel) and the 
   potential $V$ (lower panel) for the sine-Gordon KK moduli space
   with coordinates $(a,b)$. At $a=0$ there is an apparent
 boundary, reachable after a finite time.}\label{KK-sG}
 \end{figure}
\begin{figure}
 \includegraphics[width=1.0\columnwidth]{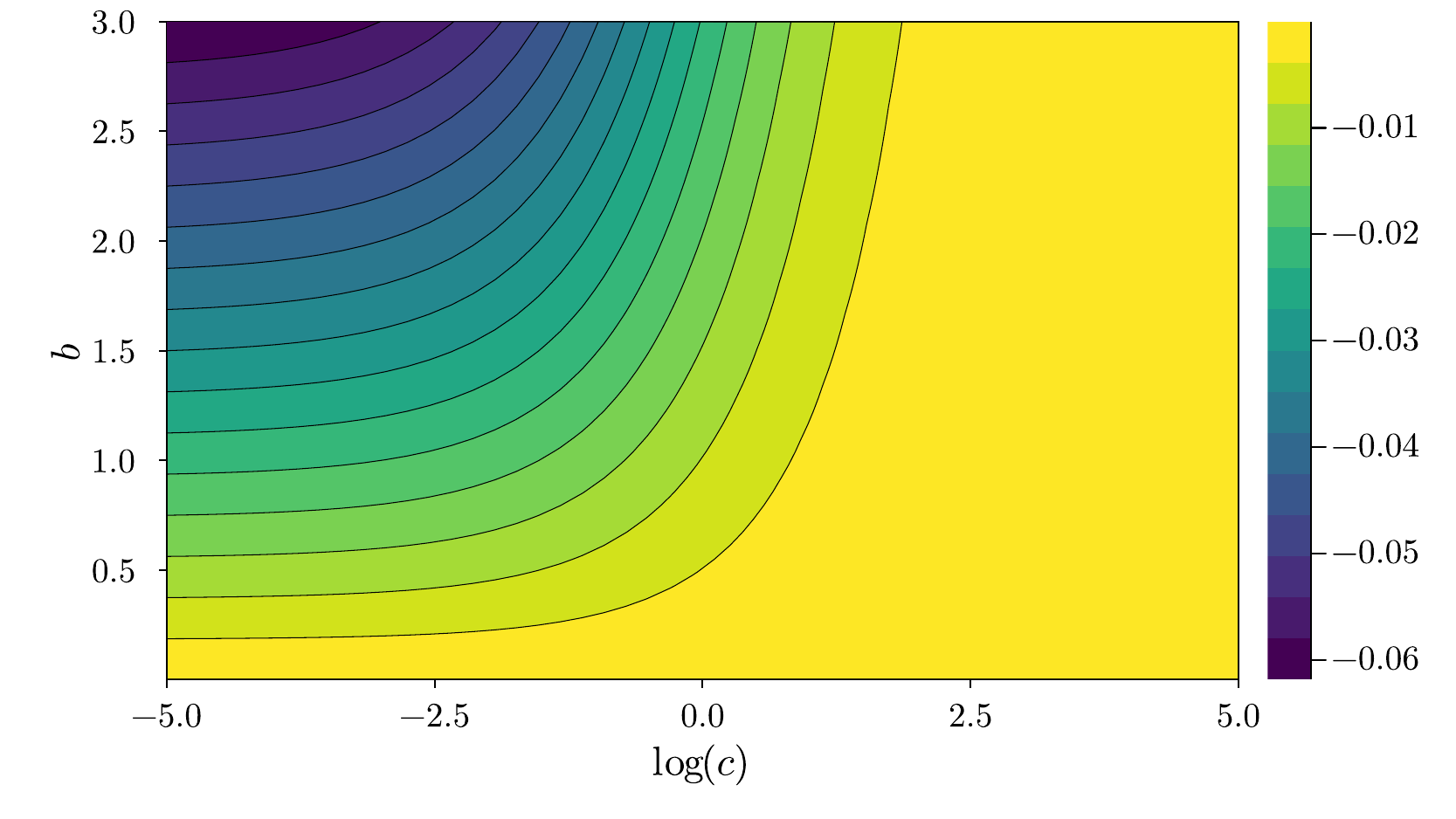}
 \includegraphics[width=1.05\columnwidth]{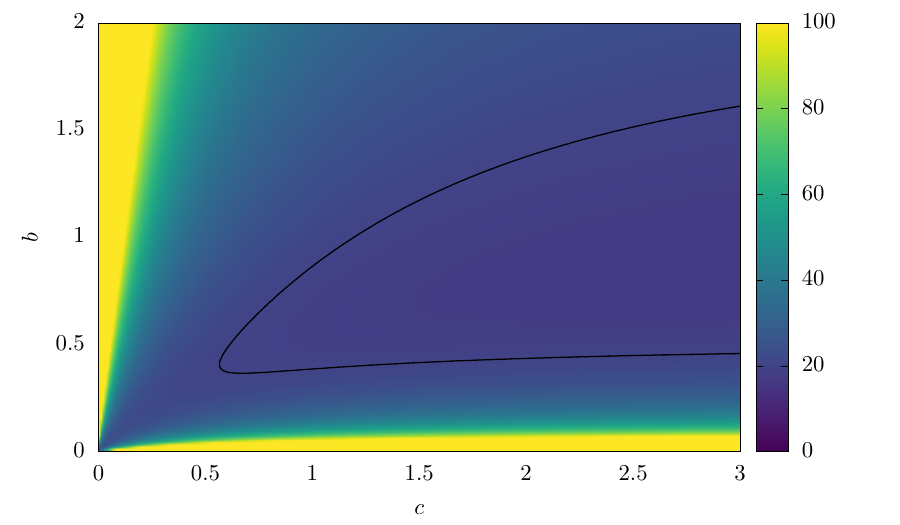}
 \caption{The Ricci curvature $R$ (upper panel) and the
   potential $V$ (lower panel) for the extended sine-Gordon
   KK moduli space with coordinates $(c,b)$. The region $c\in
   (0,1)$ was excluded in Fig. \ref{KK-sG}. At the boundary
   $c=0$ the potential diverges. }\label{KK-sG-cb}
 \end{figure}

For the KK configurations (\ref{eff-kk}), we obtain a
two-dimensional CCM of the general form
\be
L[a,b]=\frac{1}{2} g_{aa}\dot{a}^2 + g_{ab} \dot{a}\dot{b}
+\frac{1}{2} g_{bb} \dot{b}^2 - V(a,b) \,.
\ee
We find, by integration, that the metric is non-diagonal and
has components
\bea
g_{aa} &=& 16 b  \left( 1-\frac{2ab}{\sinh(2ab)} \right),\nonumber  \\
g_{ab}&=&- 16a   \frac{2ab}{\sinh(2ab)} \,, \nonumber \\
g_{bb} &=& \frac{4\pi^2}{3b^3}   \left[ 1
+ \left(1-\frac{8}{\pi^2}  (ab)^2\right)
\frac{2ab}{\sinh(2ab)} \right] ,\hspace*{0.3cm}
\eea
satisfying the identity (\ref{Identityab}),
and that the potential is
\bea
V(a,b) &=& 8 \Bigg[ b \left( 1+\frac{2ab}{\sinh(2ab)} \right) \nonumber \\
&& \quad + \frac{1}{b\tanh^2(ab)}
\left( 1-\frac{2ab}{\sinh(2ab)} \right) \Bigg] \,.
\eea
The CCM supports a simple solution
\be
b=\gamma, \;\;\; a=\frac{1}{\gamma} \mbox{arcosh}
\frac{\cosh (\gamma v t)}{v} \,,
\ee
which, when inserted into (\ref{eff-kk}), reproduces the exact
kink-kink solution (\ref{exact-kk}). In fact, this result can be
anticipated by directly comparing the exact solution and the
restricted set of configurations (\ref{eff-kk}).

Let us now have a closer look at the KK moduli space $\mathcal{M}$. As
expected, for $a \to \infty$ it models two independent
relativistic solitons, and therefore the CCM has twice the Lagrangian
(\ref{K-sG-RCCM}). On the other hand, $\mathcal{M}$ naively has a
boundary at $a=0$, because two of the metric components have zeros:
\be
g_{aa}= \frac{32b^3}{3} a^2  + O(a^4) \,, \quad g_{ab} = -16 a +O(a^3)
\,.
\ee
However, this is only an apparent singularity. Indeed, the Ricci
curvature remains finite at $a=0$, see Fig. \ref{KK-sG}. This null
vector problem, resulting from the vanishing of $\partial_a \Phi$ as
$a \to 0$, can be resolved by a more appropriate choice of
coordinates, leading not only to well behaved metric functions but
also, importantly, allowing us to extend the moduli space beyond $a=0$.

The resolution is achieved by reparametrizing
the configurations (\ref{eff-kk}) as
\be
\tilde{\Phi}_{KK}(x;c,b)= 4\arctan \frac{\sinh bx}{c} \,,
\ee
where the new collective coordinate $c$ is related to the old ones
via $c=\cosh (ba)$. The metric components are now
\bea
g_{cc}&=&\frac{16}{b} \frac{1}{c^2-1}
\left( 1-\frac{1}{c \sqrt{c^2-1}} \mbox{arcoth} \frac{c}{\sqrt{c^2-1}}
\right) \,, \nonumber  \\
g_{cb} &=& -\frac{16}{b^2} \frac{1}{\sqrt{c^2-1}}
\mbox{arcoth} \frac{c}{\sqrt{c^2-1}} \,,  \nonumber \\
g_{bb}&=& -\frac{4}{b^3} \frac{1}{\sqrt{c^2(c^2-1)}}\left(- \sqrt{c^2(c^2-1)}
\left( \ln^2(\tilde{c}) +\frac{\pi^2}{3} \right)\right. \nonumber \\
&+&\left. \frac{1}{6} \ln^3(\tilde{c}) +\frac{\pi^2}{6}
\ln(\tilde{c})\right) \,,
\label{gcc}
\eea
where 
\be
\tilde{c} = -1+2c^2 - 2\sqrt{c^2(c^2-1)} \,,
\ee
and they are regular on the line $c=1$, which corresponds to
$a=0$. Since the potential $V$ also has a finite value on the line,
we can extend the moduli space to $c<1$. The curvature and
potential extended to the interval $c\in (0,1)$ are shown in
Fig. \ref{KK-sG-cb}. This interval is accessible in a finite time if the
initial state has sufficient oscillatory energy. Interestingly, this
interval of the moduli space can also be obtained in the previous
construction, but requires imaginary values of the coordinate $a$ in
the range $(0,i\pi/2b)$.

The boundary at $c=0$ is unattainable since the potential
\bea
V(c,b) &=&  \frac{8}{bc \left(c^2-1\right)^{3/2}}
\Biggl(c \sqrt{c^2-1} \left(b^2 \left(c^2-1\right)+c^2\right)
\nonumber \\
&+& \left(b^2 \left(c^2-1\right)-c^2\right) \mbox{arcoth}
\left(\frac{c}{\sqrt{c^2-1}}\right)\Biggr)
\eea
diverges as $c\to 0$. A second line of $\mathcal{M}$ where the metric
behaves badly is $b=0$. This is also a true boundary towards which the
potential diverges, so no finite-energy KK trajectories reach it.

To conclude, the extended relativistic moduli space provides a well
defined dynamical system for KK dynamics in the sG model, as well as
reproducing the exact KK scattering solution.

\subsection{Kink-antikink solution}

We turn now to the kink-antikink (KAK) solutions of the sG model, both
the scattering solution
\be
\Phi_{KAK}(x,t)=4\arctan \frac{\sinh (\gamma v t)}{v \cosh (\gamma x)}
\ee
and the breather
\be
\Phi_B(x,t)=4\arctan \left( \frac{\sqrt{1-\omega^2}}{\omega}
\frac{\sin (\omega t)}{\cosh (\sqrt{1-\omega^2} \, x)} \right) \,,
\ee
where $0 < \omega < 1$.

These solutions can be obtained by using the KAK superposition
\bea
\Phi_{KAK}(x;a,b) &=& 4\arctan e^{b(x+a)} - 4\arctan e^{b(x-a)} \nonumber \\
&=& 4\arctan \frac{\sinh ba}{\cosh bx} \,. \label{eff-kak}
\eea
Here, the configurations are not symmetric under $a\mapsto -a$,
so $a$ can have any real value. However, there is a symmetry
under the combined transformation $a\mapsto -a$ and $b\mapsto -b$, so
it makes sense to require $b > 0$. The configuration with $b=0$ is
the vacuum configuration for any $a$, but this configuration also
occurs for $a=0$ with $b > 0$. 

The restricted set of configurations (\ref{eff-kak}) defines the KAK
moduli space $\mathcal{M}$ with metric components
\bea
g_{aa}&=&16 b\left( 1+\frac{2ab}{\sinh(2ab)} \right), \nonumber \\
g_{ab}&=&16  a \frac{2ab}{\sinh(2ab)} \,, \nonumber \\
g_{bb} &=&  \frac{4\pi^2}{3b^3}   \left[ 1 - \left(1-\frac{8}{\pi^2}
(ab)^2\right) \frac{2ab}{\sinh(2ab)} \right], \hspace*{0.3cm}
\eea
and potential
\bea
V(a,b) = &8& \left[b \left( 1- \frac{2ab}{\sinh(2ab)} \right) \right.
\nonumber \\
&+&\left. \frac{\tanh^2(ab)}{b}  \left( 1+\frac{2ab}{\sinh(2ab)} \right)
\right] \,.
\eea
The identity (\ref{Identityab}) is again satisfied.
Contrary to the KK case, the potential is finite, see Fig. \ref{KAK-sG}.
However, for $a \ne 0$, the configurations with $b \to 0$ are rather far
from the physical regime $b \approx 1$, so are unimportant
for KAK scattering.

The relativistic CCM has the solution
\be
b=\gamma \,, \quad a(t)=\frac{1}{\gamma} \mbox{arsinh}
\frac{\sinh (\gamma v t)}{v} \,,
\label{sGexactscatt}
\ee
fully reproducing the analytical KAK scattering solution for all
velocities. As before, it can be obtained by comparing the exact
solution with the restricted configurations (\ref{eff-kak}). For the
breather solution,
\bea
b &=& \sqrt{1-\omega^2} \,, \nonumber \\
a(t) &=& \frac{1}{\sqrt{1-\omega^2}}
\mbox{arsinh} \left( \frac{\sqrt{1-\omega^2}}{\omega}
\sin (\omega t) \right).
\label{sGexactbreath}
\eea

\subsection{KK and KAK moduli spaces compared}

\begin{figure}
\includegraphics[width=1.0\columnwidth]{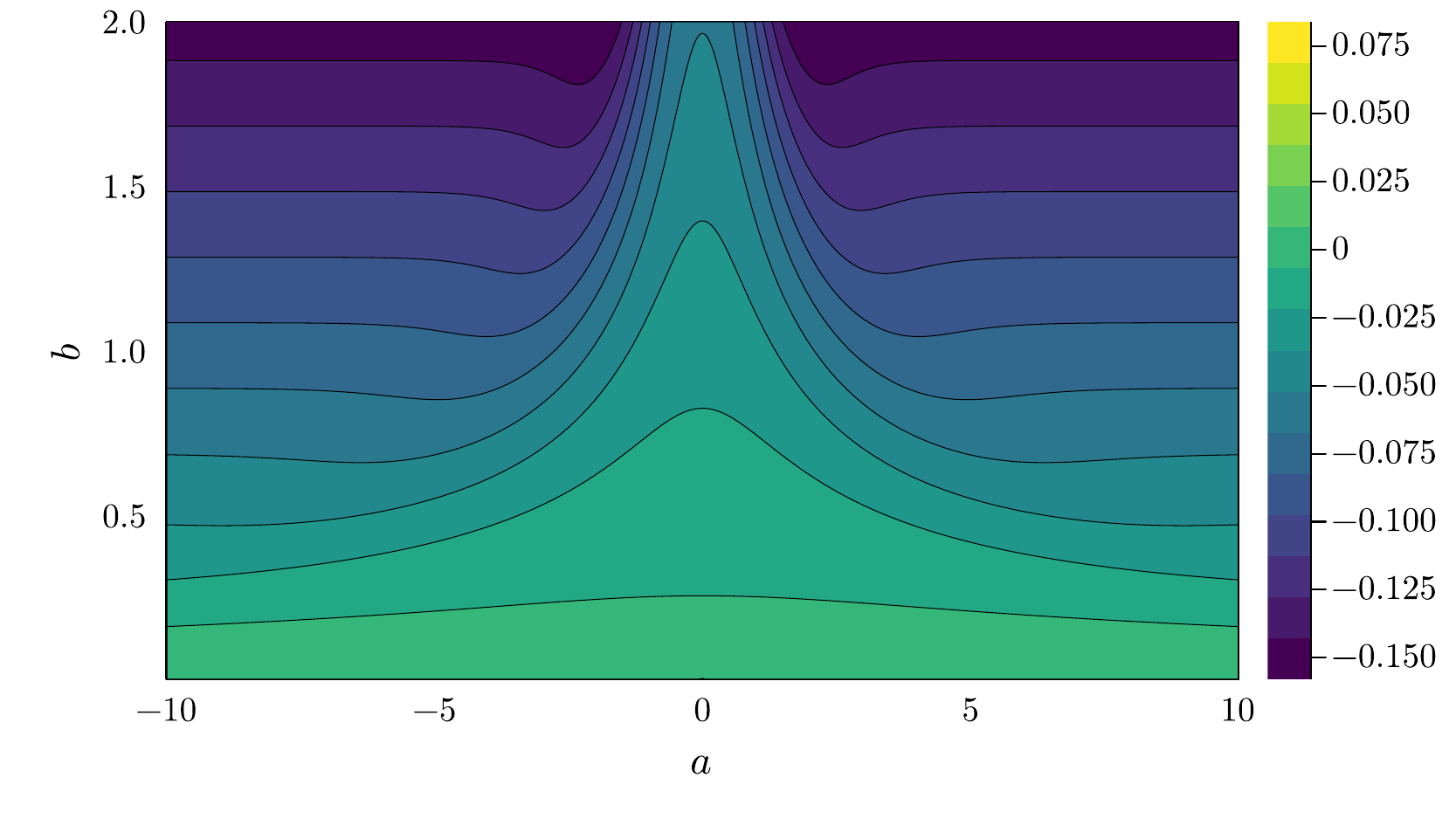}
\includegraphics[width=1.0\columnwidth]{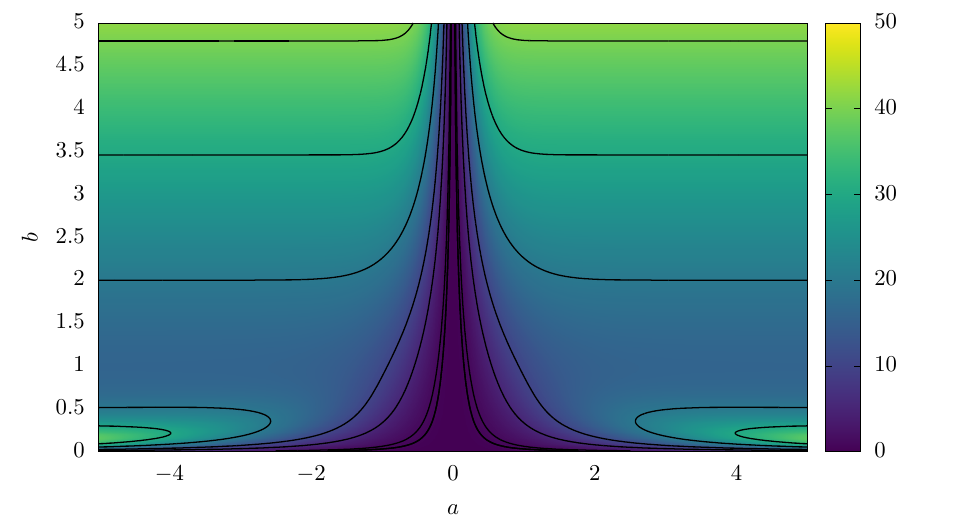}
 \caption{The Ricci curvature $R$ (upper panel) and the
   potential $V$ (lower panel) for the KAK moduli space
   in the sine-Gordon model. The boundary at $b=0$, containing a
   possible singularity, is not included.}\label{KAK-sG}
\end{figure}

The KAK moduli space has metric components $g_{ab}$ and
$g_{bb}$ whose behaviour for small $a$ is
\bea
g_{ab}&=&16 a +O(a^3) \,, \nonumber \\
g_{bb}&=& \frac{8}{9b} (12+\pi^2) a^2 +O(a^4) \,.
\eea
These components vanish at $a=0$, producing a metric singularity,
whereas it was $g_{aa}$ and $g_{ab}$ that
vanished at $a=0$ in the KK case. The Ricci curvature is still finite
for all $a$ and $b$, see Fig. \ref{KAK-sG}.
This suggests the KK and KAK sectors have different types of null vector. To
understand this better, we expand the restricted configurations
$\Phi_{KK}$ and $\Phi_{KAK}$ for small $a$, finding
\bea
\Phi_{KK}(x;a,b) &=& 4\arctan(\sinh(bx)) \\
&& \quad -\frac{2\sinh(bx)}{1+\sinh^2(bx)}b^2a^2 + O(a^4) \nonumber
\eea
and
\be
\Phi_{KAK}(x;a,b) = \frac{4}{\cosh(bx)} ba +O(a^3) \,.
\ee
For $\Phi_{KK}$ the derivative w.r.t. $a$ vanishes as $a\to 0$, while for
$\Phi_{KAK}$ it is the derivative w.r.t. $b$. In the former case it
was sufficient to introduce the coordinate
$c = \cosh(ba) \approx 1 + \frac{1}{2}b^2a^2$, but
this does not help in the latter case, where the singularity is
essential, as we argued in Section III.

We recall here that there exists a relativistic model of two interacting point
particles on the line, whose dynamics precisely reproduces the
kink-kink (or kink-antikink) solution of the sine-Gordon model. This
is the famous Ruijsenaars--Schneider model with a particular pair
potential \cite{RS}. Undoubtedly, it would be very interesting to
understand possible relations between this model and the  relativistic
CCM. 

\subsection{Perturbative framework applied to sG model}

Here we set $b = 1 + \epsilon$ and expand in $\epsilon$.
For the single sG kink we find 
\bea
&& \Phi_K(x;a,b) = 4\arctan e^{b(x-a)} =
4\arctan e^{(x-a)}  \\ 
&& \qquad +\epsilon\frac{2(x-a)}{\cosh(x-a)}
-\epsilon^2\frac{(x-a)^2 \tanh(x-a)}{\cosh(x-a)}
+ o(\epsilon^2) \,. \nonumber
\eea
The set of configurations up to first order is therefore
\be
\Phi_K(x;a,C)  = 
4\arctan e^{(x-a)} + C \frac{2(x-a)}{\cosh(x-a)} \label{pert-K}
\ee
where $C \equiv C_1$. Their two-dimensional moduli space has a
diagonal metric with components
\bea
g_{aa} &=& 8+8C+\frac{2}{9}(12+\pi^2)C^2 \,, \nonumber \\
g_{CC} &=& \frac{2\pi^2}{3} \,.
\eea
The quadratic character of $g_{aa}$ makes this a wormhole-type metric.
\begin{figure*}
\includegraphics[width=\textwidth]{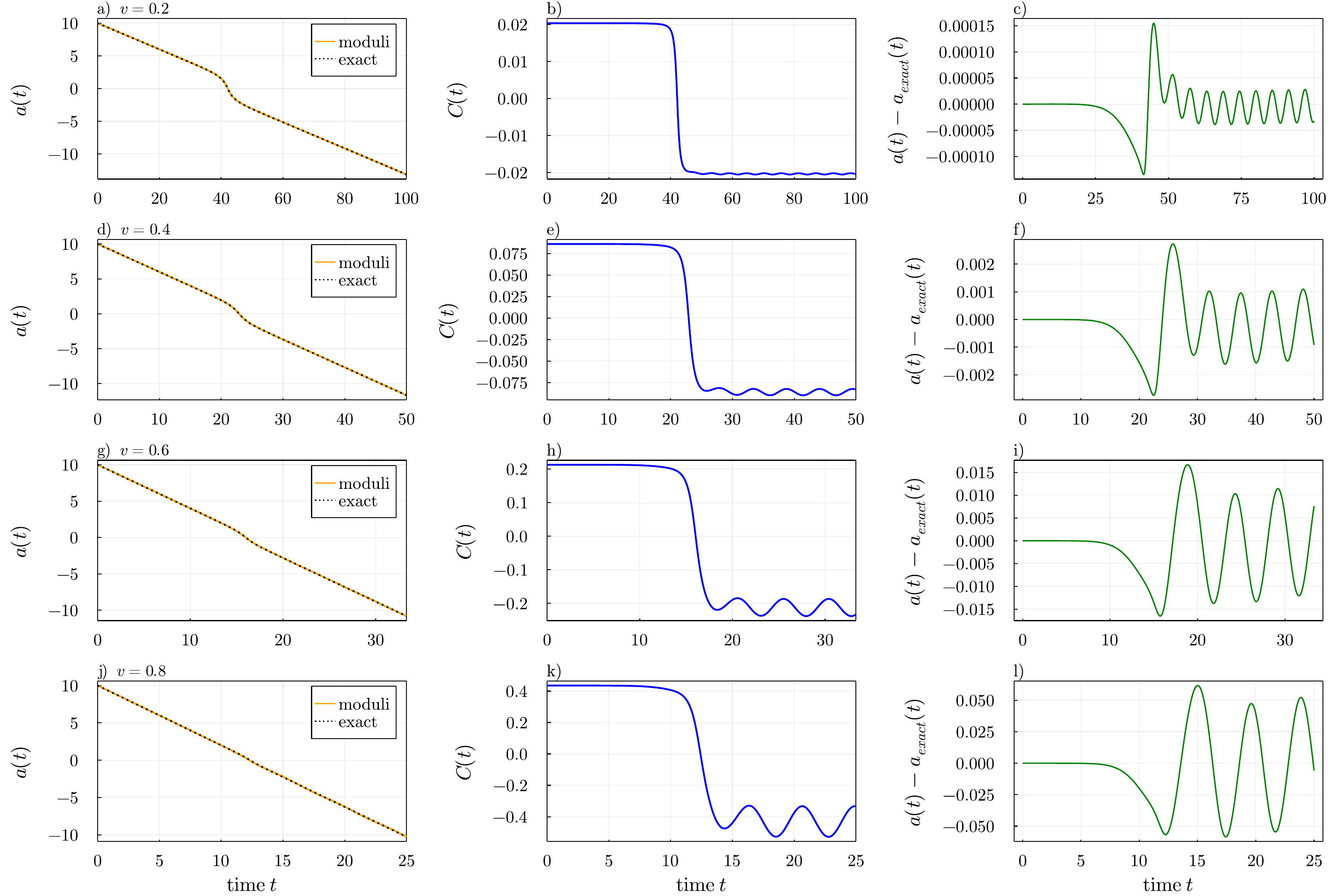}
\caption{KAK collision in the pRCCM for the sine-Gordon model with
  moduli $(a,C)$.
  Left: $a(t)$; Center: $C(t)$;
  Right: Difference between $a(t)$ obtained in the pRCCM and the exact
  expression $a_{exact}(t)$. From top to bottom
  $v=0.2, 0.4, 0.6$, and $0.8$.}
   \label{sG-aC}
\end{figure*}
\begin{figure}
 \hspace*{-0.2cm} \includegraphics[width=1.05\columnwidth]{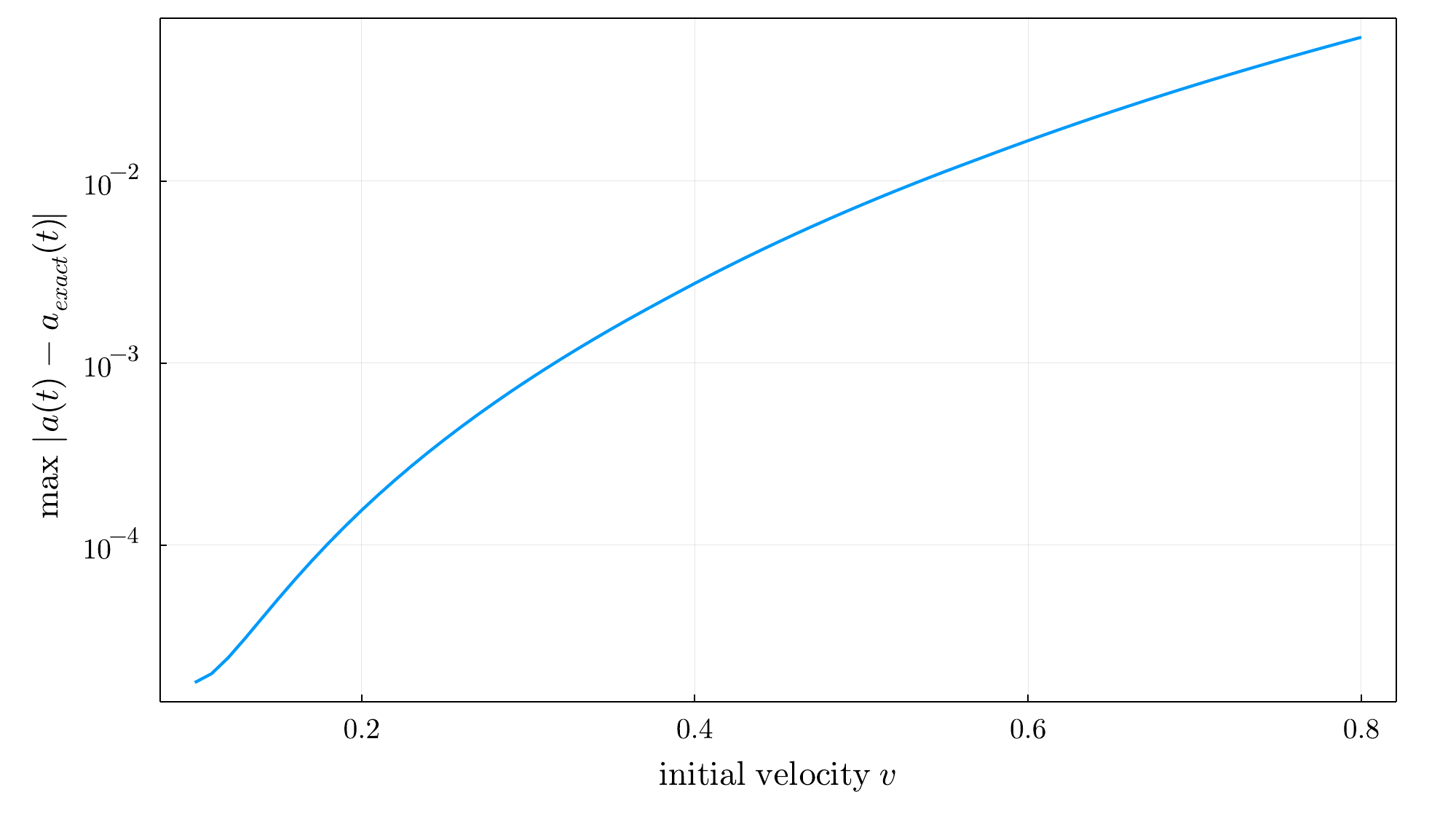}
  \caption{Maximum of the difference between $a(t)$ obtained in the
    pRCCM for the sine-Gordon model and the exact expression
    $a_{exact}(t)$, as a function of $v$.}
   \label{sG-a-aex}
  \end{figure}  

The potential, up to fourth order in $C$, is
\bea
V(C) &=& 8+ 4C^2 +\frac{2}{9} \left(\pi ^2-6\right) C^3 \nonumber \\
   &+& \frac{1}{450} \left(240-100 \pi ^2+7 \pi ^4\right) C^4 \,.
\eea
Importantly, the equations of motion of the resulting pRCCM have a stationary
solution satisfying
\bea
\ddot{a} &=& 0 \,, \label{zeroaccel} \\
\frac{1}{2} \partial_C g_{aa} \, {\dot{a}}^2  &=& \partial_C V \,.
\label{stationary-eq}
\eea
The first equation has the constant-velocity solution $\dot{a}=v$, while
the second becomes the nonlinear algebraic equation
\bea
0 &=& 4 v^2 + \left(\frac{2}{9} \left(12+\pi^2\right) v^2-8\right)C \\
&-&\frac{2}{3} \left(\pi^2-6\right) C^2
-\frac{2}{225} \left(240-100 \pi^2+7 \pi^4\right)C^3 \nonumber \,,
\eea
with a real solution whose truncated expansion is
\be
{\tilde C} \approx 0.5 v^2 + 0.223127 v^4
 + 0.072637 v^6 + 0.016753 v^8 \,.  \label{stat-sol-phi}
\ee
This expansion reproduces the true value of $C$ with good precision.

\begin{figure*}
\includegraphics[width=\textwidth]{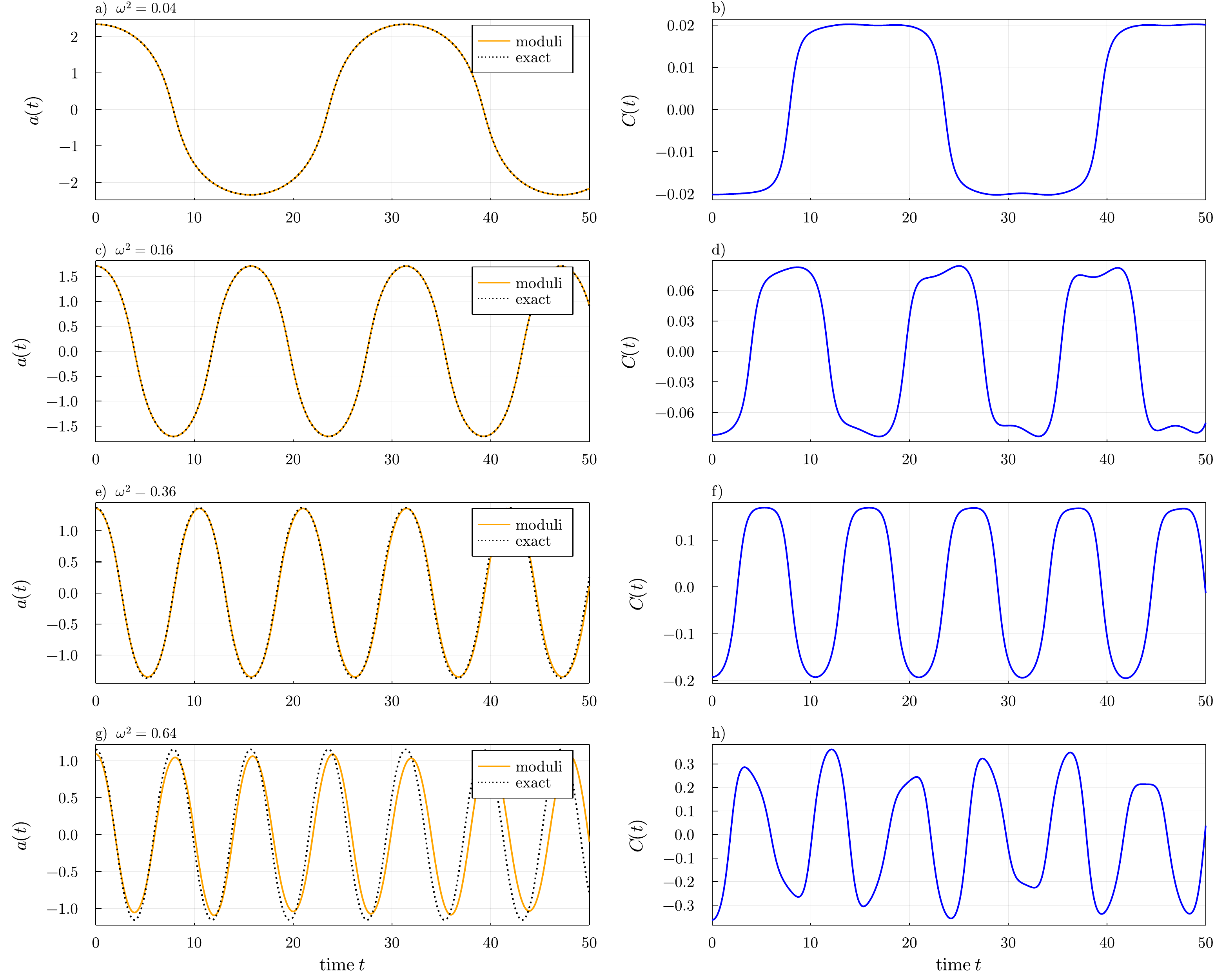}
\caption{Oscillatory solutions $a(t)$ (left) and $C(t)$ (right) of the
pRCCM for the sine-Gordon model. Here (from top to bottom)
$\omega^2=0.04,0.16,0.36$, and $0.64$. The dotted line represents
$a(t)$ computed from the exact breather solution.}
   \label{sG-aC-br}
\end{figure*}

Obviously by enlarging the model, taking into account more terms
in the expansion, we approach the exact Lorentz-boosted kink with arbitrary
accuracy. The price is the growing dimension of the moduli space and
the complexity of the resulting equations.

Although the perturbative expansion seems to be an unnecessary
complication for a single kink, its usefulness becomes clear when
we apply it to the KAK solutions in sG. We
consider only the terms up to first order, using
the moduli $(a,C)$. The restricted set of
configurations, where the null vector is already removed, is
\bea
&& \Phi_{KAK}(x;a,C) =  4\arctan e^{(x+a)} - 4\arctan e^{(x-a)} \nonumber \\
&& \quad + \frac{2C}{\tanh(a)}\left( \frac{x+a}{\cosh(x+a)}
-\frac{x-a}{\cosh(x-a)} \right) \,. \label{pKAK-sG}
\eea
The resulting metric is non-singular for any finite $(a,C)$, but as
the formulae for the metric components are very long, we do not present
them. Furthermore, in our numerical investigations of KAK dynamics
we obtain the metric and potential numerically, not referring to
their analytical formulae (see Appendix A). In this way, we arrive at
a well defined two-dimensional pRCCM for modelling KAK collisions in
the underlying field theory. This procedure could be repeated up to
any order $n$.

We also need appropriate initial conditions for the approaching kink
and antikink. These are provided by the moving single-kink solution of
the pRCCM with $C$ given by (\ref{stat-sol-phi}), which leads to the
following initial conditions at $t=0$,
\bea
a(0)&=&a_0, \;\; \dot{a}(0)=v, \nonumber \\
C(0)&=&0.5 v^2 + 0.223127 v^4 + 0.072637 v^6 + 0.016753 v^8,
\nonumber \\
\dot{C}(0)&=&0 \,, \label{sG-init}
\eea
where $a_0$ is half the initial distance between the solitons.

In Fig. \ref{sG-aC}, left and center column, we present the time evolution
of $a(t)$ and $C(t)$ obtained in the pRCCM, based on the configurations
(\ref{pKAK-sG}) and the initial conditions (\ref{sG-init}).
The initial velocity has the range of values $v=0.2,
0.4, 0.6$, and $0.8$. For comparison, we also plot the position of the
kink (and antikink) obtained from the sG KAK solution $a_{exact}(t)$
given in (\ref{sGexactscatt}). 

To better see the agreement between the pRCCM computation and the
exact solution, we plot the difference $a(t)-a_{exact}(t)$ in
Fig. \ref{sG-aC}, right column. The difference is small even for
quite relativistic velocities. E.g., for $v=0.4$ the maximum difference is
about $0.003$, while at large times it oscillates between $\pm
0.001$.

In Fig. \ref{sG-a-aex} we plot the maximum of this difference as a
function of the initial velocity $v$. For small velocities, where
our first-order relativistic approximation should work well, the
agreement is extremely good. E.g., for $v \lesssim 0.2$ the difference
is less than $10^{-4}$. This should be compared with the
one-dimensional moduli space computation where for $v=0.2$ the
corresponding difference is of order $10^{-2}$ \cite{zero-vec}. Hence,
the inclusion of the first Derrick mode leads to results
which are two orders of magnitude more precise.

It is also important to notice that the inclusion of the Derrick
mode does not spoil the integrability of the model. The result of the
KAK scattering is always a KAK pair. We do not observe any bounce
windows or bion formation. This is obviously a crucial test for the
validity of our perturbative relativistic framework.

The same pRCCM can be used to study the breather solution, provided
the initial conditions have energy less than twice the
static energy of a kink. In Fig. \ref{sG-aC-br} we present examples of
oscillatory solutions $(a(t), C(t))$. The
coordinate $a(t)$ is also compared with the exact expression
(\ref{sGexactbreath}) derived from the breather solution. We find good
agreement for $\omega \ll 1$. For larger frequencies we notice some
discrepancy. There is also a departure from exact periodicity. In
any case, it is an important result in our framework that the addition
of the new collective coordinate does not lead to the disappearance of
the bound orbits. It should be underlined that in the
one-dimensional CCM \cite{zero-vec} the existence of periodic
solutions is guaranteed if the energy condition for a breather is
satisfied. In higher-dimensional moduli spaces, on the other hand, it
is a nontrivial phenomenon, which relies on the details of the moduli space
metric and potential.

\section{$\phi^4$ theory}

We consider next the prototypical, non-integrable scalar
field theory supporting topological solitons, the $\phi^4$ theory in
(1+1)-dimensional space-time. The Lagrangian is
\be
L_{\phi^4}[\phi]=\int_{-\infty}^\infty dx \left( \frac{1}{2} (\pa_t\phi)^2
  - \frac{1}{2}(\pa_x\phi)^2 - \frac{1}{2} (1-\phi^2)^2\right) \,,
\ee
and the field equation has the static BPS kink solutions 
\be
\Phi_K(x;a) = \tanh (x-a) \label{kink}
\ee
interpolating between the vacua $-1$ and $+1$. The modulus
$a \in \mathbb{R}$ is the position of the kink, and the resulting
canonical moduli space has the constant
metric $g_{aa}=M=4/3$, so the CCM Lagrangian is
\be
L[a]=\frac{2}{3} \dot{a}^2 -\frac{4}{3} \,.
\ee
In this description, the kink moves at an arbitrary
velocity $v$ as a non-relativistic particle with mass $M$ and the
Lorentz invariance of the original field theory is lost.
The antikink $\Phi(x;a) = -\tanh (x-a)$ interpolates between
$+1$ and $-1$, but otherwise has similar properties to the kink.

The $\phi^4$ kink has one positive-frequency normal mode,
the normalized {\it shape mode}
\be
\eta_{sh}(x)=\frac{3}{2}\frac{\sinh(x)}{\cosh^2(x)} \,,
\ee
whose frequency $\omega^2_{sh}=3$ is below the continuum threshold
$\omega^2 = 4$. This mode plays a distinguished role in multi-kink
dynamics in $\phi^4$ theory.

\subsection{Relativistic Moduli Space}

Including the Derrick scaling deformation, with modulus $b>0$, the
single-kink configurations are 
\be
\Phi_K(x;a,b) = \tanh b(x-a) \,.
\label{boost}
\ee
The resulting two-dimensional moduli space has the diagonal metric
\be
g_{aa}=\frac{4}{3} b \,, \quad g_{bb}= \frac{\pi^2-6}{9} \frac{1}{b^3}
\,,
\ee
and potential
\be
V(b)= \frac{2}{3} \left(b +\frac{1}{b} \right) \,,
\ee
so the relativistic CCM combining these has Lagrangian
\be
L[a,b]=\frac{2}{3}b \dot{a}^2 + \frac{\pi^2-6}{18b^3} \dot{b}^2
- \frac{2}{3} \left(b +\frac{1}{b} \right) \,.
\ee
As before, the equations of motion have solutions describing
relativistic, Lorentz-contracted kinks moving at constant velocity.

The normalized Derrick mode is now
\be
\eta_D(x)=\frac{3}{\sqrt{\pi^2-6}} \frac{x}{\cosh^2(x)} \,,
\ee
which is known to be almost identical to the shape
mode \cite{MORW}. Indeed, the Derrick mode frequency
$\omega_D^2=12/(\pi^2-6) \approx 3.101$ is very close to the shape
mode frequency, and the inner product of the normalized Derrick and
shape modes $(\eta_D,\eta_{sh})\approx 0.98$ is very close to unity. 

Exactly as in the sine-Gordon case, the construction of a relativistic
CCM describing KAK collisions encounters difficulties. The
superposition of kink and antikink, 
\be
\Phi_{KAK}(x;a,b) = \tanh b(x+a) - \tanh b(x-a) - 1 \,,
\label{KAKconfigs}
\ee
leads to a CCM with metric components  
\bea
g_{aa} &=& b \left[ \frac{8}{\sinh^2(2ab)} \left( -1
+ \frac{2ab}{\tanh(2ab)} \right) +\frac{8}{3} \right], 
\\
g_{ab} &=& \frac{8a}{\sinh^2(2ab)}
\left( -1 + \frac{2ab}{\tanh(2ab)} \right), \nonumber
\\
 g_{bb} &=& \frac{2}{9b^3}  \Biggl[ \pi^2 - 6 \nonumber \\
&& + \frac{3}{\sinh^2(2ab)} \left( \pi^2 - \frac{2ab}{\tanh(2ab)}
(\pi^2 - 8a^2 b^2) \right) \Biggr],  \nonumber
\eea
and potential 
\bea
V(a,b)&=& \frac{4b}{3} \Biggl[\frac{3}{\sinh^2(2ab)}
\left(1-\frac{2ab}{\tanh(2ab)} \right)+1 \Biggr] \nonumber \\
&+& \frac{4}{3b} \Biggl[ -17-24ab -\frac{3(5+12 ab)}{\sinh^2(2ab)} \\
&+& \frac{6}{\tanh(2ab)}\left( 3+4ab +\frac{5ab}{\sinh^2(2ab)}
\right) \Biggr] \,. \nonumber
\eea
The metric components satisfy (\ref{Identityab}), and once again the Ricci
curvature is finite for any $a\in \mathbb{R}$ and $b > 0$, but the metric
still degenerates at $a=0$. There is a null vector problem at $a=0$, because
the field configurations (\ref{KAKconfigs}) become the vacuum, which has
zero derivative w.r.t. $b$. However, as argued earlier, there is no simple
resolution of this singularity. The configuration space for $a > 0$ is
a smooth two-dimensional manifold, and similarly for $a < 0$. These two
surfaces are glued together at the single point $a=0$, and the result is
singular.

If we introduce the coordinate $c=ab$ in (\ref{KAKconfigs}), then
\be
\Phi_{KAK}(x;b,c) = \tanh (bx+c) - \tanh (bx-c) -1 \,.
\ee
The corresponding metric components on the moduli space are 
\bea
{\tilde g}_{bb} &=& \frac{2}{9b^3} \Biggl[ 12 c^2+\pi^2 -6 \nonumber \\
&+&  \frac{3}{\sinh^2(2c)} \left(12 c^2 +\pi^2 -\frac{2c}{\tanh(2c)}
\left(4 c^2+\pi ^2\right) \right) \Biggr] \,, \nonumber \\
{\tilde g}_{bc} &=&  -\frac{8}{3} \frac{c}{b^2} \,, \nonumber \\
{\tilde g}_{cc} &=& \frac{1}{b}  \left[ \frac{8}{\sinh^2(2c)}
\left( -1 + \frac{2c}{\tanh(2c)} \right) +\frac{8}{3} \right] \,,
\eea
which satisfy the identity (\ref{Identitybc}), and the potential $V$
simplifies when expressed in terms of $b$ and $c$.

\subsection{Perturbative approach to KAK collisions}

To avoid the singularity at $a=0$, we use the perturbative approach.
As before, we begin with the Derrick deformed single kink
$\Phi_K = \tanh((1+\epsilon)(x-a))$ with modulus $b=1+\epsilon$,
and perform an expansion in $\epsilon$. If we keep only the first
Derrick mode $\eta_D$, then
\be
\Phi_K(x;a,C) = \tanh (x-a) + C \frac{x-a}{\cosh^2(x-a)} \,,
\ee
where we have identified $C=\epsilon$. These configurations have a
moduli space with the diagonal metric 
\bea
g_{aa}&=& \frac{4}{3} + \frac{4}{3}C + \frac{4\pi^2}{45}  C^2 \,, \nonumber \\
g_{CC} &=&\frac{\pi^2-6}{9}  \,,  \label{phi4-metric}
\eea
and the potential
\bea
V(C) &=& \frac{4}{3} + \frac{2}{3}C^2
+ \frac{1}{45} (4 \pi^2-30)C^3 \nonumber \\
&+& \frac{1}{450} (120-70 \pi^2+6 \pi^4)C^4 \,.
\eea
As in the case of the sG kink, the pRCCM has a stationary solution
obeying the equations of motion (\ref{zeroaccel}) and
(\ref{stationary-eq}). Specifically, the equation for $C$ reads
\bea
0&=&\frac{2 v^2}{3} +\frac{4}{45} (\pi ^2
   v^2-15)C -\frac{1}{15} (4 \pi^2-30)C^2 \nonumber \\
&-&\frac{2}{225} (120-70 \pi^2+6 \pi^4)C^3 \,,
\eea
when we set $\dot{a}=v$. This can be solved exactly, but as the
solution has a long, unilluminating form, we use its truncated
expansion in $v^2$, 
\be
{\tilde C} \approx 0.5 v^2 + 0.210507 v^4+0.027428v^6-0.0302351v^8 \,,
\ee
in our numerical simulations.

Keeping the first two Derrick modes, the single-kink configurations are
\bea
\Phi_K(x; a,{\bf C}) &=&\tanh(x-a)+ C_1 \frac{x-a}{\cosh^2(x-a)} \nonumber \\
&-& C_2 (x-a)^2 \frac{\tanh (x-a)}{\cosh^2(x-a)} \,,
\eea
which leads to a pRCCM with three moduli,
\be
L[a,{\bf C}] = \frac{1}{2}g_{aa} \dot{a}^2 +
\frac{1}{2} g_{ij} \dot{C}_i \dot{C}_j - V({\bf C}) \,.
\ee
Here we explicitly use the fact that $g_{ai}=0$, $i=1,2$. Although the
metric functions and the potential can be found analytically, the
formulae are again long, so we do not display them. In any case, in our
calculations they are computed numerically.       
There is a stationary solution approximating the boosted kink moving
at constant velocity $\dot{a}=v$, having non-zero Derrick mode amplitudes
$\tilde{C}_k$ determined using the algebraic equations
\be
\frac{1}{2} \pa_{C_k} g_{aa} v^2 = \pa_{C_k} V \,.
\ee

We turn now to kink-antikink (KAK) collisions, and compare the
results from the pRCCM keeping either one or two Derrick moduli to
results from full field theory simulations, and also to the results
found in ref.\cite{MORW} using the CCM based on the shape mode.
\begin{figure*}
 \includegraphics[width=1.03\columnwidth]{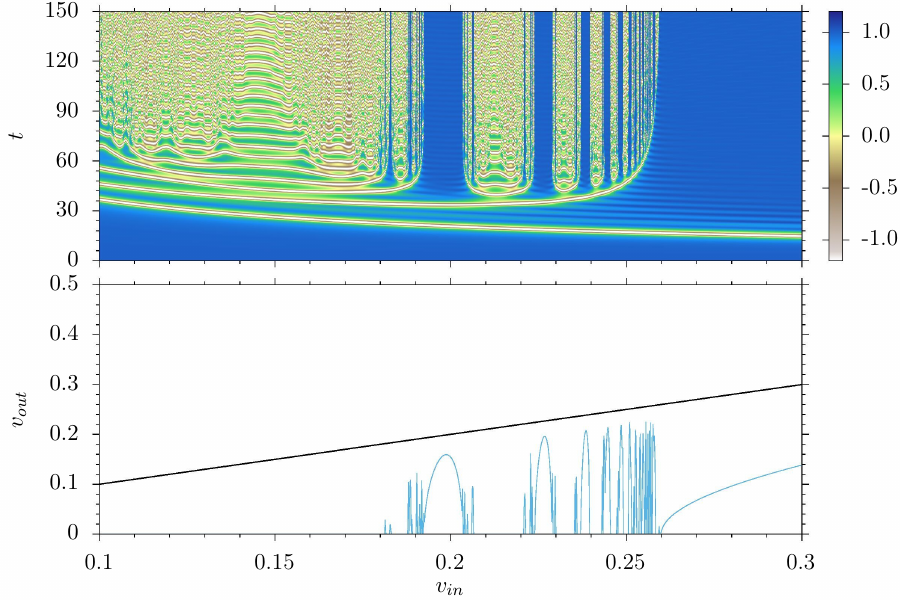}
 \includegraphics[width=1.03\columnwidth]{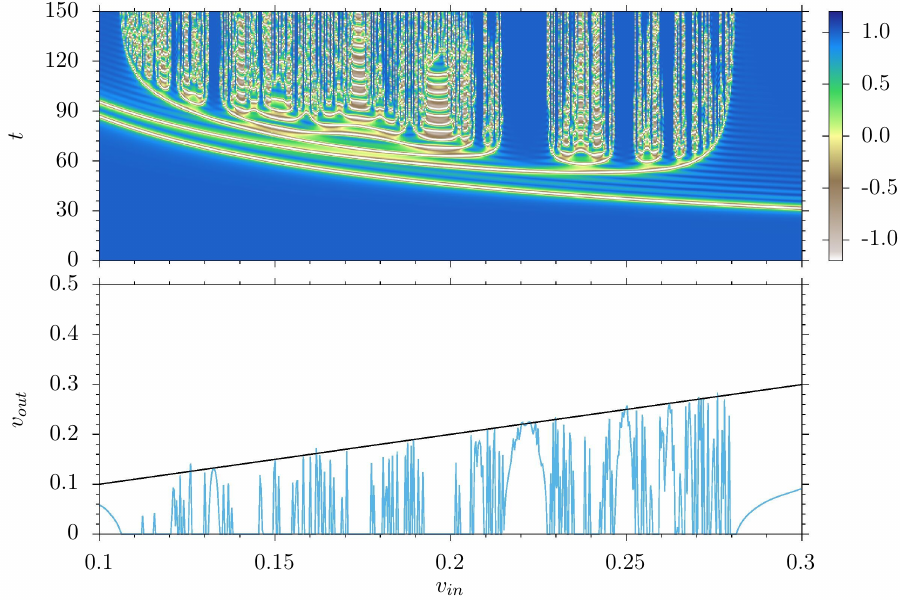}
 \includegraphics[width=1.0\columnwidth]{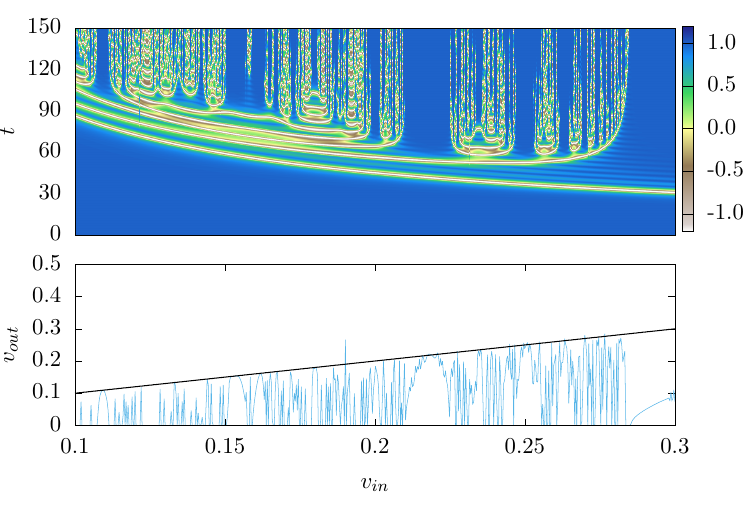}
 \includegraphics[width=1.0\columnwidth]{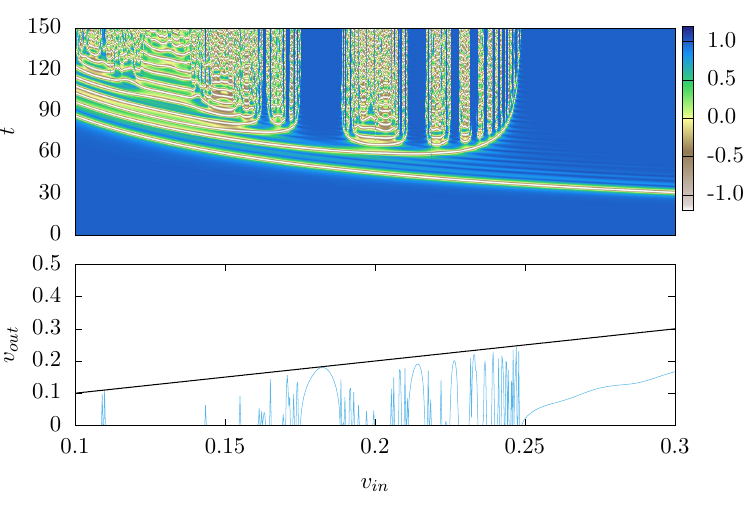}
\caption{KAK collision in $\phi^4$ theory:  time dependence of the
  field at the origin, $\phi(x=0,t)$, and final velocity of
  backscattered antikink $v_{out}$ for various initial velocities
  $v_{in}$. {\it Upper left:} the full field theory dynamics;
  {\it upper right:} the CCM based on the shape mode \cite{MORW};
  {\it  lower left:} the pRCCM with moduli $(a,C)$;
  {\it lower right:} the pRCCM with moduli $(a,C_1,C_2)$.} 
\label{Fig-phi4}
\end{figure*}

The configurations in the pRCCM are kink-antikink superpositions expanded
to finite order in the Derrick modes,
\be
\Phi_{KAK}(x;a,{\bf C})=\Phi_K(x;-a,{\bf C}) - \Phi_K(x;a,{\bf C})
-1 \,.
\ee
Inserting these into the $\phi^4$ theory Lagrangian gives
\be
L_{KAK}(a,{\bf C}) = \int_{-\infty}^\infty
\mathcal{L}_{\phi^4} [\Phi_{KAK}(x;a,{\bf C})] \, dx \,,
\ee
and after integrating over $x$ we obtain the Lagrangian of the pRCCM,
which is a well defined dynamical system provided the
null-vector problems are cured by redefining the moduli via 
\be
C_k \mapsto \frac{C_k}{\tanh(a)} \,.
\ee 
The resulting second-order equations of motion for the moduli
require initial conditions 
\bea
&&a(0)=a_0, \;\; \dot{a}(0)=v \,, \nonumber \\
&&C_k(0) =\tilde{C}_k, \;\; \dot{C} (0)=0 \,,
\eea
corresponding to a well-separated KAK pair boosted towards each other.

More explicitly, the configurations modelling KAK collisions with just one
Derrick mode are
\bea
&& \Phi_{KAK}(x;a,C) = \tanh(x+a) - \tanh(x-a) -1 \nonumber  \\
\hspace*{-0.4cm} &+&
\hspace*{-0.0cm} \frac{C}{\tanh(a)} \left( \frac{x+a}{\cosh^2(x+a)}
-\frac{x-a}{\cosh^2(x-a)} \right) \,. \label{pKAK-phi4} 
\eea 
The resulting pRCCM has equations of motion that must be supplemented
by the single-kink initial conditions
\bea
a(0)&=&a_0 \,, \quad \dot{a}(0)=v \,, \nonumber \\
C(0)&=&0.5 v^2 + 0.210507 v^4+0.027428v^6-0.0302351v^8 \,, \nonumber \\ 
\dot{C}(0) &=& 0 \,.  \label{phi4-init}
\eea

In the full $\phi^4$ field theory simulations of KAK collisions,
the kink and antikink are boosted towards each other at speeds $v_{in}$,
through the initial conditions
\bea
&& \phi_{in}(x,0) =  
\tanh(\gamma (x+a_0)) - \tanh(\gamma(x-a_0)) - 1 \,, \nonumber \\
&& \partial_t \phi_{in}(x,0) =  \frac{\gamma v_{in}}
{\cosh^2(\gamma (x+a_0))} - \frac{\gamma v_{in}}{\cosh^2(\gamma (x+a_0))} \,.
\nonumber \\
&&
\eea

In the field theory the kink and antikink either perform a few
{\it bounces} before escaping to infinity, or they annihilate to the
vacuum via the formation of an oscillating state, called a {\it bion},
which slowly radiates, decaying to the vacuum. Interestingly, these two
possibilities exhibit an amazingly complicated pattern depending on
the incoming soliton velocities, with multi-bounce
windows and bion chimneys occurring in a fractal manner, see
Fig. \ref{Fig-phi4}, upper left panel. The fractal structure starts at
$v_{min}\approx0.18$ and ends at $v_{crit}= 0.2598$. Below $v_{min}$
only bion chimneys exist, while above $v_{crit}$ only one-bounce
scattering is observed.

For a comparison with the non-relativistic CCM based on the shape
mode \cite{MORW}, see Fig. \ref{Fig-phi4}, upper right panel. The
fractal structure is {\it qualitatively} reproduced but there are
important details which do not fully agree: (i) there is an overall
shift toward larger values of $v_{in}$, e.g., $v_{crit} \approx 0.28$;
(ii) there is an unwanted, wide two-bounce window that dominates the
low-velocity dynamics for $v_{in} \lesssim 0.11$; (iii) there are
many three-bounce and higher-bounce windows in the field theory's
bion regime, i.e., for $0.11 \lesssim v_{in} \lesssim 0.2$. Of course,
the appearance of bounce windows
with a large number of bounces is unsurprising as the CCM has no
radiation modes that could transfer energy from the bion. In the CCM,
bions can decay only to a free KAK pair. However, the existence of
additional two-, three- or four-bounce windows is a rather unwanted
feature. Despite this, the results provide convincing evidence
that resonant energy transfer between kink motion and an oscillatory mode
is the mechanism responsible for the observed fractal structure in
the final state formation.

The dynamics found in the pRCCM with only one Derrick modulus $C$ looks
similar, see  Fig. \ref{Fig-phi4}, lower left panel. The overall shift
to higher velocities persists, e.g., $v_{crit}=0.2853$. The velocity
shift compared with the field theory is $\Delta v_{crit} = 0.0255$,
a $10\%$ error. However, there is an important improvement in the
low-velocity regime, as the number of unwanted low-velocity bounce
windows is drastically reduced. Specifically, there is no low-velocity
two-bounce window and there is only one unwanted three-bounce window. 

A spectacular improvement is observed if we use the pRCCM keeping two
Derrick moduli $C_1$ and $C_2$, see  Fig. \ref{Fig-phi4}, lower right
panel. Basically, almost all bions in the bion chimneys behave as in
the field theory, i.e., they do not decay into free solitons after
just a small number of bounces. Hence, there are very few bounce windows in
the low-velocity regime, and each exhibits a large number
of bounces. A similar improvement is observed in the higher-velocity
regime. Furthermore, the critical velocity is substantially reduced, to
$v_{crit} = 0.2490$, for which $\Delta v_{crit}= 0.0108$, which is
only a $4\%$ error. Additionally, the small wiggles in
the final velocities of the solitons, occurring in both the
two-dimensional CCMs, are now absent.

Interestingly, the frequency of the first Derrick mode tends to the
frequency of the shape mode as the dimension of the pRCCM
increases. Specifically, it decreases from $\omega^2_1=3.1011$ to
$\omega^2_1=3.0221$ when the modulus $C_2$ is included. In the latter
case, the second Derrick mode has frequency $\omega^2_2=6.9283$,
which is above the continuum threshold. Therefore, oscillations of
this mode may partially simulate some aspects of radiation.

\section{Conclusions}

In the present work we have explored the relativistic collective coordinate
model for (multi-)kink dynamics, which arises when the Derrick scaling
deformation with modulus (scale parameter) $b>0$ is included.
For a moving single kink the model reproduces the Lorentz
contraction of the kink \cite{Rice}, and its reduced
Lagrangian is that of a relativistic point particle.

The model can be extended to kink-kink (KK) and kink-antikink (KAK)
collisions. For the KK sector of the sine-Gordon model, we have constructed a
globally well defined moduli space whose metric and potential
lead to a Lagrangian reproducing the exact KK solution.

In the case of general KAK dynamics, the construction encounters some
difficulties. The moduli space has a metric singularity at $a=0$, i.e., when
the kink and antikink coincide and pass through the
vacuum solution. We have shown that this is an essential geometric
singularity that cannot be removed by a redefinition of the collective
coordinates. 

To circumvent this issue, we have introduced a perturbative version
of the relativistic collective coordinate model (pRCCM), where the
Derrick modulus is expanded around its undeformed value $b=1$. This
corresponds to an expansion in the squared kink velocity, and
therefore incorporates
relativistic corrections in a perturbative manner. Here, a key idea
is to treat all the terms in the field expansion as {\it independent}
higher-order Derrick modes whose amplitudes are independent
collective coordinates. The field configurations are less constrained
than before, even if just the first-order (original) Derrick mode
is retained. Each of these new collective coordinates has a null
vector problem at $a=0$, but these can all be resolved by a coordinate
redefinition that absorbs a factor $\tanh(a)$. Note that such a
coordinate redefinition has no physical effect for $a>0$, but it smoothly
extends the moduli space through $a=0$ and allows for a smooth dynamics. 

In contrast to previous treatments of the singularity of the
relativistic CCM, this is a straightforward approach that can easily
be implemented. Furthermore, it provides an arbitrary number of collective
coordinates, which can be used to improve the description of  
multi-kink collisions in any (1+1)-dimensional scalar field
theory. It has been tested in the sG model and in $\phi^4$ theory, and
the results are very encouraging.

In the case of KAK collisions in the sG model, we found that the
pRCCM does not spoil the integrability property. The
inclusion of the first Derrick mode gives a model where there is
always a separating kink and antikink in the final state. There
is no annihilation or particle production, as expected.
In fact, the two-dimensional pRCCM gives a more accurate
approximation, by two orders of magnitude, than the simpler
one-dimensional CCM constructed from a superposition of an undeformed
kink and antikink.

The pRCCM framework, applied to KAK collisions in $\phi^4$ theory,
reproduces the fractal structure of the velocity-dependence of the final
state formation rather well. When the first-order Derrick mode is included,
there is a small improvement in comparison with the CCM incorporating
the kink's shape mode \cite{MORW}. The improvement is particularly
striking when both the first-order and second-order Derrick modes are
included. The results from the pRCCM appear to be converging to those of
the field theory as the number of retained Derrick modes increases.

This shows the universality of the Derrick deformation framework.
Namely, it works for qualitatively distinct kinks, one having a
shape mode, and the other not. It would be interesting to extend this
relativistic framework to field theories with kinks that are
not spatially antisymmetric, e.g., to the kinks of the $\phi^6$ model. 

Finally, one would like to extend the framework to higher
dimensions. For example, in the context of (3+1)-dimensional
Skyrmions, it may improve the vibrational quantization
procedure \cite{chris}. This is, of course, related to the question of
possible quantum corrections to kink dynamics, recently reconsidered
by Evslin \cite{Jarah}. Quantum corrections can contribute to shape
mode dynamics, and therefore presumably to the dynamics of Derrick
modes at any order.

\section*{Acknowledgements}

CA and AW acknowledge financial support from the Ministry of Education, 
Culture, and Sports, Spain (Grant FPA2017-83814-P), the Xunta de 
Galicia (Grant INCITE09.296.035PR and Conselleria de Educacion) and 
the Spanish Consolider Program Ingenio 2010 CPAN (Grant CSD2007-00042). 
This work has received further financial support from Xunta de Galicia 
(Centro singular de investigación de Galicia accreditation 2019-2022), from 
the European Union ERDF, and from the “María de Maeztu” Units of Excellence 
program MDM-2016-0692 and the Spanish Research State Agency.
KO, TR and AW were supported by the Polish National Science Centre 
(Grant NCN 2019/35/B/ST2/00059). We thank Jose Queiruga for calling
our attention to ref. \cite{Rice}, and Maciej Dunajski for pointing out an
error in the $\phi^4$-theory KAK metric in an earlier version of this
paper. AW also thanks Zoltan Bajnok and Romuald Janik.  

\vskip 20pt

\appendix

\section{Numerical method for solving the CCM} 
\label{a2}

The collective coordinate model for field configurations $\Phi(x;{\bf X})$
with Lagrangian (\ref{eff-lag}) has the Euler--Lagrange equations
\bea
\hspace{-0.5cm} \ddot X^i \int_{-\infty}^\infty
\frac{\partial\Phi}{\partial X^i}\frac{\partial\Phi}{\partial X^k} \, dx 
&+& \dot X^i\dot X^j  \int_{-\infty}^\infty
\frac{\partial^2\Phi}{\partial X^i\partial X^j}
\frac{\partial\Phi}{\partial X^k} \, dx \nonumber \\ 
&=& \quad -\int_{-\infty}^\infty \frac{\partial W }{\partial X^k} \, dx
\eea 
where
\be
W(\Phi)= \frac{1}{2} \left( \frac{\partial \Phi}{\partial x} \right)^2
+ U(\Phi) \,.
\ee
This can be rewritten as a matrix differential equation for the moduli $X^i$,
\bea
\left( \int_{-\infty}^\infty e_ie_k \, dx \right)\ddot X^i &+&
\left(\int_{-\infty}^\infty H_{ij}e_k \, dx \right)\dot X^i\dot X^j
\nonumber \\
&& \quad = -\int_{-\infty}^\infty \frac{\partial W}{\partial X^k} \, dx
\eea
where
\be
e_i=\frac{\partial\Phi}{\partial X^i} \,, \qquad
H_{ij}=\frac{\partial^2\Phi}{\partial X^i\partial X^j} \,.
\ee

In solving this set of equations, it turns out that
calculating the integrals numerically, even for the metric integrand $e_ie_k$,
is more stable than implementing analytical expressions. The cost is
similar to the cost of calculating the integral on the right hand
side, as we need to numerically calculate the vectors $e_i$
anyway. Direct analytical calculations are prone to many numerical
errors, arising, for example, from the catastrophic cancellation
problem. Therefore, we calculated each required integral at every time step.

\end{document}